\documentclass[twocolumn,aps,pre,floats,showpacs]{revtex4}
\newcommand{\beq}{\begin{eqnarray}}
\newcommand{\eeq}{\end{eqnarray}}

\usepackage{epsfig}% Include figure files
\usepackage{dcolumn}% Align table columns on decimal point
\usepackage{bm}% bold math
\usepackage{amsfonts}
\usepackage{amsmath}
\usepackage{amssymb}
\usepackage{latexsym}
\usepackage{pdfsync}

\def\be{\begin{equation}}
\def\ee{\end{equation}}

\def\d{\partial}
\def\rf#1{(\ref{#1})}
\def\rfs#1{Eq.~\rf{#1}}

\begin{document}

\title{Non-adiabacity and large flucutations in a many particle Landau Zener problem}

% \author{Alexander Altland$^1$, V. Gurarie$^2$, T. Kriecherbauer$^3$,
% A. Polkovnikov$^4$}
\author{Alexander Altland$^1$}
\author{V. Gurarie$^2$}
\author{T. Kriecherbauer$^3$}
\author{A. Polkovnikov$^4$}

\address{$^1$Institut f\"ur theoretische Physik, Z\"ulpicher Str.~77,
  50937 K\"oln, Germany}
 \address{$^2$Department of Physics, CB390, University of Colorado,
   Boulder CO 80309, USA}
 \address{$^3$Fakult\"at f\"ur Mathematik, Ruhr-Universit\"at Bochum,
   Bochum, Germany}
 \address{$^4$Department of Physics, Boston University, Boston MA 02215, USA}

\date{\today}
%\maketitle
%\parbox{14cm}
\begin{abstract}
  We consider the behavior of an interacting many particle
  system under slow external driving -- a many body generalization of
  the Landau-Zener paradigm. We find that a conspiracy of interactions
  and driving leads to physics profoundly different from that of the
  single particle limit: 
  for practically all values of the driving rate the particle
  distributions in Hilbert space are very broad, a phenomenon caused
  by a strong amplification of quantum fluctuations in
  the driving process. These fluctuations are 'non-adiabatic' in that
  even at very slow driving it is exceedingly difficult to push the
  center of the distribution towards the limit of full ground state
  occupancy. We obtain these  results  by a number of
  complementary theoretical approaches, including diagrammatic
  perturbation theory, semiclassical analysis, and exact
  diagonalization.
\end{abstract}
 \maketitle
% 03.65.Sq Semiclassical theories and applications
% 03.65.Yz Decoherence; open systems; quantum statistical methods
% 05.45.Mt Quantum Chaos; semiclassical methods

%\bcols
%\maketitle
%\begin{multicols}{2}
\bigskip

\section{Introduction}

In many physical contexts, one is met with quantum systems that are
subject to `slow' time dependent external driving. The most elementary
prototype in this category, the Landau-Zener (LZ) system
\cite{Landau1932,Zener1932}, contains just two coupled levels driven linearly
in time. In this system, the initially occupied instantaneous ground
state level of its Hamiltonian
\begin{equation}
  \label{eq:2}
  \hat H =
  \left(
    \begin{matrix}
      \lambda t&  g \cr
      g  &-\lambda t
    \end{matrix}
  \right),
\end{equation}
stays occupied in the infinite future with a probability
\be
P = 1-e^{-\frac{\pi g^2}{\lambda}}.
\label{lz}
\ee where $g$ is the coupling strength, $\lambda$ defines the driving
rate and $\exp(\pi g^2/\lambda)$ is the so-called Landau-Zener
parameter. The approach $P\stackrel{\lambda\to 0}{\longrightarrow} 1$
is manifestation of the quantum adiabatic theorem, i.e. the statement
that sufficiently slow driving keeps a quantum system in its adiabatic
ground state.

Many quantum single particle systems can be effectively described in
terms of the Landau-Zener setup, or one of its multi-dimensional
generalizations \cite{Demkov1966,Osherov1966,Kayanuma1985,Brundobler1992}. The reason is that the driven approach of pairs of
instantaneous eigenstates will generate 'avoided crossings' which can
be represented by Hamiltonians such as in (\ref{eq:2}). The cumulative
statistics of these crossings then describes the behavior of the
system in the course of the driving process. In particular, the system
will remain in its ground state if only the latter is sufficiently
well separated from the first excited states.

But how do many particle systems behave under driving? Given the
exponential abundance of energy levels  in interacting
systems (or the fact that superimposed on the groundstate we often
have an continuum of low lying, 'soft' excitations) reference to the
adiabatic theorem will not be sufficient to understand the
consequences of driving.  In some instances, it is possible to reduce
the problem to one of studying the statistics of linear (oscillator)
excitations superimposed on an invariant ground state. A system of
this type has been studied in work by Yurovsky, Ben-Reuven and Juliene \cite{Yurovsky2002} (see also \cite{Kayali2003}), with the
principal observation that the driving process generates a number of
$x$ excitations, where
\begin{equation}
  \label{eq:1}
  x\equiv \exp(\pi g^2/\lambda)
\end{equation}
coincides with the Landau-Zener parameter.

In general, however, many particle systems cannot be linearized.  The
ramifications of non-linearities in a driven context have been studied
in Ref.~\cite{Niu2000,Niu2002}, within the simplifying framework of a
low-dimensional system (specifically, a Bose-condensed system which
can be described in terms of a single complex condensate amplitude.)
Describing its dynamics in terms of a nonlinear Schr\"odinger
equation, Ref.~\cite{Niu2002} observed examples of rather interesting
behavior, including situations where the system does not remain in the
ground state even in the fully adiabatic limit. (For this, however,
authors had to consider systems of bosons with attractive
interactions. In practice, such systems tend to collapse.) The problem
is also particular in that it has a low
dimensional Hilbert space.

% {\em Recently there was an increased activity in analyzing slow
%   dynamics in low-dimensional critical systems~\cite{ap_05, zurek_05,
%     dziarmaga_05, sengupta_08, dutta, fazio}. In all these cases it
%   was argued that the number of excitations produced at slow rates
%   scales as a power law of the rate (as opposed to the the exponential
%   dependence Eq.~(\ref{lz}) for the Landau-Zener problem). The
%   exponent of this power law is related to the critical exponents
%   characterizing this phase transition~\cite{ap_05,
%     zurek_05}. Furthermore in Ref.~\cite{np} it was argued that the
%   power law dependence of various thermodynamic observables is generic
%   for all gapless systems. In low dimensions with high-density of low
%   energy states there is a stronger tendency to have smaller power,
%   i.e. be in a less adiabatic regime. In systems with bosonic
%   excitations there can be additional tendency to be less adiabatic
%   related to bosonic bunching~\cite{np}. In this work using a
%   particular model (see below) we show that a similar power law
%   behavior of number of created excitations with the rate can be
%   understood in terms of classical adiabatic invariants. These
%   invariants are usually conserved at slow processes~\cite{LL1}, which
%   is equivalent to having no transitions between energy levels in
%   quantum systems. However, near singularities, where the period of
%   the motion in a classical orbit diverges, the adiabatic invariants
%   significantly change in time leading to non-analytic power law
%   dependence of various observables with the rate.}

Refs.~\cite{Altland2008,Pazy2006} considered an interacting
driven system which is generic in that it shows the two principal
characteristics of interacting quantum systems: a high dimensional
Hilbert space and nonlinearity. In one of its representations, the
system describes a large assembly of degenerate fermions which may
pair-combine into a bosonic level as their energy gets pushed up (a
cartoon of a fermion/boson conversion process as realized in, say, a
time-dependent BEC/BCS crossover.)  Quite strikingly, it turned out
that this system resists getting close to its adiabatic ground
state, i.e. a state where all particles eventually have become
bosonic; even at very slow driving an $\mathcal{O}(1)$ fraction of
particles remains in energetically high-lying sectors of Hilbert
space. In fact, the methods employed in Ref.~\cite{Altland2008} did
not  enable us to get close to the 'adiabatic regime' of the
model. The approach to adiabacity was discussed in Ref.~\cite{Pazy2006} by mapping
the evolution of the system to an effective
Hamiltonian dynamical system. In this way, as we show in this paper,
the exponential dependence (\ref{lz}) gives way to a power law,
$$
1-{n_b\over N} \sim \lambda,
$$
controlling the ground state occupancy for a wide range of initial
conditions.  Here $n_b\in [0,N]$ is the number of particles in the
bosonic ground state, and $N$ the total number of particles.
Formally, the reluctance of the system to approach its ground state
may be understood in terms of dynamical instabilities~\cite{Pazy2006}
(to be discussed in some detail further below.) Physically, it
reflects the general inertia of interacting systems to adjust to (time
dependent) environmental changes.

In this paper, we will discuss an interaction phenomenon which is no
less remarkable: the slow dynamical evolution of the system goes along
with an exceptionally strong buildup of quantum fluctuations: although
we are dealing with a system that should behave 'semiclassically', on
account of the largeness of its number of particles, $N$, the
distributions of particles in Hilbert space turn out to be very broad.
This phenomenon can be attributed to an amplification  of
quantum fluctuations: classically, the driving process changes the
relative energy of two many particle systems. At some point, the
originally lower (stable) system becomes higher in energy
(unstable). In a strict classical sense, however, the ground state of
the elevated system remains stationary. It takes the action of weak
($\mathcal{O}(N^{-1})$) quantum fluctuations to destabilize this state
and initiate the evolution towards energetically more favorable
sectors of Hilbert space. The nonlinearity of that evolution
(interactions) then leads to an amplification of the initial
fluctuations, up to a point where they become of
$\mathcal{O}(1)$. Specifically, we find that
\begin{itemize}
% \item In order to increase the ratio of the occupation of the ground
%   state, $n_b$, to the total number of particles, $N$, to a value
%   $n_b/N = 1- \epsilon$, the driving rate has to decrease as
%   \begin{equation}
%     \label{eq:3}
%     \lambda \sim  { \epsilon \pi g^2\over \ln(N)}.
%   \end{equation}
%   Specifically, a full conversion of all but $\mathcal{O}(N^0)$
%   particles to bosons -- the adiabatic limit -- calls for driving
%   rates of $\mathcal{O}(N^{-1})$, and exceedingly small
%   Landau-Zener parameters $x=\mathcal{O}(\exp(N))$.  This is to be
%   compared to the case of single particle systems (with discrete
%   spectrum) where $\lambda \lesssim \pi g^2$ or $x=\mathcal{O}(1)$ is
%   sufficient to reach adiabacity.
\item in all but the extreme adiabatic limit $N-n_b\ll N$ the ground
  state occupation $n_b$ shows massive fluctuations,
  $\mathrm{var}(n_b)/ \langle n_b\rangle^2 = \mathcal{O}(1)$. At
  intermediate mean occupancy $\langle n_b\rangle \simeq N/2$, the
  distribution $P(n_b)$ extends over almost the full interval $[0,N]$.
\item At fast rates $n_b\ll N$, the probability distribution is exponential:
  $P(n_b)\propto \exp[-n_b/\langle n_b\rangle]$. At slower rates, it
  becomes even broader and covers Hilbert space in a manner for which
  we have no analytic expressions. At yet slower rates, upon
  approaching the adiabatic limit, the distribution gets
  'squeezed' into the boundary region $n_b/N\sim 1$. It then assumes a
  universal Gumbel form~\cite{Gumbel}, the latter frequently appearing
  in the context of extreme value statistics.
\end{itemize}
In this paper, we will set the stage for the discussion of
fluctuations by establishing contact between our earlier 'quantum'
approach to the problem -- the latter adjusted to regimes of fast and
moderate driving -- and a slow driving semiclassical formulation. Our
semiclassical approach is different from that 
Ref.~\cite{Pazy2006} in that it microscopically connects to the early
quantum stages of the evolution, a matching procedure
necessary to explore the role of fluctuations. (In
fact, we also doubt validity of the results for the \textit{mean}
conversion rates derived in that earlier
reference: for all but zero initial occupancy of the boson state, a
power law in the driving parameter is predicted that we believe is
incorrect.)

The rest of the paper is organized as follows: in section
\ref{sec:model} we will introduce the spin variant of the model, and
discuss the mapping to its other representations. We will then analyse
the system, in a manner that is structured according to the driving
rate: in section \ref{sec:holst-prim-regime} we discuss the regime of
moderately high driving rates, where the system can be effectively
linearized and admits a full analytic solution. In section
\ref{sec:keldysh} we go beyond the linear regime and consider driving
rates $1< \lambda^{-1}< \mathcal{O}(\ln N)$ in terms of effective rate
equations. Although these equations become uncontrolled for values
$\lambda^{-1}\sim \mathcal{O}(\ln N)$, they are good enough to signal
the system's reluctance to approach the adiabatic limit. In
section~\ref{sec:twa} we will formulate the semiclassical approach to
the model, and the so-called truncated Wigner approximation. This
approximation becomes highly accurate at sufficiently large $N\gtrsim
10$, and for all values of the driving rate. In section
\ref{sec:deepadiabaticlimit} we use the method of adiabatic invariants
to explore the semiclassical theory at very slow driving,
$\lambda^{-1} =\mathcal{O}(\ln N)$.  In section \ref{sec:numerics},
the quality of the results obtained in this way   will be checked by
comparison to direct numerical solutions of the Schr\"odinger equation
of our problem (which is feasible for particle numbers up to $N =
\mathcal{O}(10^3)$) and to the simulations of the semiclassical
approximation (the latter extensible  to much larger $N$.) We will
also compare to previous work in the literature.
In section \ref{sec:alt} we discuss a number of ramifications of our
problem relating to previous theoretical and experimental work. We
conclude in section \ref{sec:conclusions}.

\section{The model}
\label{sec:model}

In this section we will introduce the theoretical model discussed in
much of the rest of the paper -- a high-dimensional generalization
of the spin-boson model. We will also introduce a number of less
abstract equivalent representations, some resonant with concrete experimental
activity.

\subsection{Definition of the model}
\label{sec:definition-model}

Consider an $\mathrm{SU}(2)$ spin of value $S=N/2\gg 1$, coupled to a
time varying magnetic field of strength $(-\lambda t)$ in
$z$-direction. This will be the first quantum system participating in
the driving process. Its partner system is a single bosonic mode at
energy $(-\lambda t)$. We couple these two compounds by declaring that
the creation of a boson goes along with a lowering of the spin by
one. The total system is then described by the Hamiltonian
\begin{equation}
  \label{eq:6}
  \hat H = -  \lambda t ~ b^\dagger  b + \lambda t S^z +
  \frac{g}{\sqrt{N}}  \left(  b^\dagger \,
    S^- +  b \, S^+ \right),
\end{equation}
where $g/\sqrt N$ defines the coupling strength and $S^\pm =S_x \pm i
S_y$. The Hamiltonian (\ref{eq:6}) obeys the conservation law $[\hat
H, S^z+b^\dagger b]=0$, showing that the total Hilbert space
dimensionality of the problem is $2S+1=N+1$. The linear growth of the
dimension of Hilbert space in $N$ is of course not representative for
'generic' interacting systems (dimensionality exponential in $N$.) An
increase in the dimensionality of the problem can be effected by
symmetry breaking, e.g., by replacement of the large spin by an
assembly of $N$ non-degenerate spin $1/2$ compounds (cf. discussion in
section \ref{sec:altern-model-repr} below.) While this generalization
will make the problem largely un-tractable, we believe it to have
little qualitative effect, as long as the band-splitting is smaller
than the hybridization strength with the boson mode.

Below it will be useful to think of $b$ as a transverse magnetic field
acting on the spin. The above conservation then implies that the
transverse field strength is proportional to the deviation of the spin
off total polarization $S_z=S$. This feedback mechanism of the spin
precession into the field strength encapsulates the effect of
interactions in the spin variant of our model.  However, before
proceeding, let us introduce a few alternate representations which
make the interpretation of the model as one of interacting particles
more transparent (cf. Fig. \ref{fig:model}):
\begin{enumerate}
\item The variant dominantly discussed in earlier work describes the
  hybridization of energetically degenerate pairs of spinful fermions
  with a boson mode -- cf. the Hamiltonian (\ref{eq:4}) below. This
  may be viewed as a dispersionless approximation of the fermion-boson
  conversion processes realized in BEC-BCS crossover experiments in
  fermionic condensates ~\cite{Jin2004, Ketterle2004},
  cf. Fig.~\ref{fig:model}, b). (Although the initial full occupancy
  of the flat fermion band assumed in AG may not be adequate to the
  description of the experimental situation~\cite{Sun:2008qm}, see
  section \ref{sec:alt} below.)
\item Identifying the empty (doubly occupied) configurations of the
  fermion levels with the two states of a spin 1/2, the system maps
  onto a time-dependent variant of the Dicke
  model~\cite{Dicke:1954fc}. In this form it is relevant to the
  description of super radiance phenomena in molecular
  magnetism~\cite{Chudnovsky:2002ss}, and to cavity QED with many two
  level systems~\cite{Raimond2001}.
\item In a somewhat less obvious representation,
  the model describes the conversion of pairs of bosons into
  dimers. In this incarnation it is of relevance to recent
  experimental work of the JILA group. In these experiments,
  identical~\cite{Hodby:2005uf} or different~\cite{Papp:2006fx}
  species of atoms undergo sweep through a Feshbach resonance to form
  di-atomic molecules. (On the level of effective classical equations
  of motions, this correspondence was noted in earlier
  references\cite{Pazy2006}. Below, we will establish the connection
  within a fully quantum mechanical setting.)
\end{enumerate}
For the convenience of interested readers, the equivalence between
these different incarnations of the model is established in subsection
\ref{sec:altern-model-repr} below.  While the `spin-boson' formulation
(\ref{eq:6}) does not relate to concrete physical systems in an
obvious way, we find it ideally suited to our theoretical analysis and
much of our later discussion will be formulated in this
language. However, it is straightforward to transcribe all conclusions
to the context of the other models.

\subsection{Formulation of the problem}

The problem we will address in much of the paper is formulated as
follows. Start the dynamics in the distant past in the adiabatic
ground state of the problem: spin fully polarized $S_z =S=N/2$ and
zero bosons $n_b(t) \equiv \langle b^\dagger(t) b(t)\rangle = 0$. The
goal is to find the number of produced bosons at large positive times,
\begin{equation}
  \label{eq:7}
  n_b \equiv \lim_{t\to \infty} n_b(t)\equiv \lim_{t\to \infty}
  \langle b^\dagger(t) b(t)\rangle.
\end{equation}
The adiabatic limit is reached when $n_b \to N$ or, equivalently,
$S_z=-S$. Alternatively, we can speak of the representation in terms
of bosonic atoms and bosonic molecules, given by Eqs.~\rf{eq:17} and
\rf{eq:hamdifbos}, where this initial condition implies the absence of
atoms in the distant past, and the goal is to find the number of
produced atoms at large positive times.
%%%
%%%

As was anticipated in the introduction, the operator $b^\dagger b$
exhibits strong quantum fluctuations. It will, thus, be of interest to
explore the \textit{distribution} $P(n)$ defined by the moments
$\langle (b^\dagger b)^k\rangle$. Also, there may be physical
situations where it is not appropriate to start the process in the
spin-up polarized state. We have already mentioned the case of dilute
fermion gases~\cite{Sun:2008qm} where a near south polar configuration
may be a more appropriate starting point. Similarly, a downward
polarized state $S_z=-S$ (in an initially downward pointing field)
will model an atom $\to$ molecule conversion process with a purely
atomic starting configuration.

\begin{figure}[h]
  \centering
  \includegraphics[height=2.in]{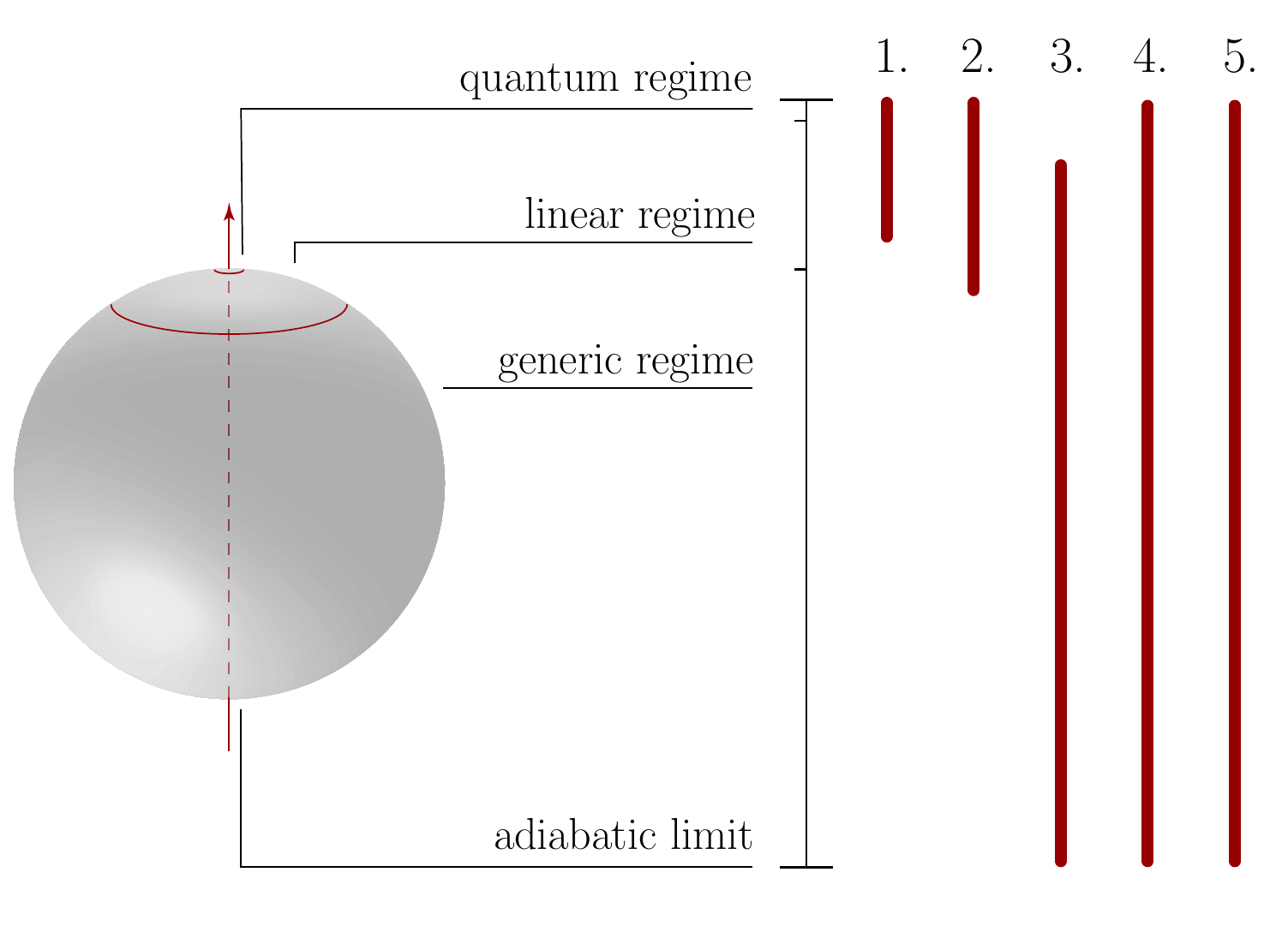}
  \caption{\label{figures} Different theoretical approaches to the
    problem. The Hilbert space of the problem contains a number of
    different sectors: a (quantum fluctuation) dominated region where
    only a few bosons have been created $n_b=\mathcal{O}(1)$ and the
    spin remains nearly polarized, a 'linear region', where the number
    of bosons may be large but is small enough to justify
    linearization of spin operators, $1\ll n_b \ll N$, a 'generic'
    region, $n_b = \mathcal{O}(N)$, and the adiabatic limit $n_b \to
    N$. Theoretical approaches we will use to approach these regimes
    include, 1.) solution of the linearized Schr\"odinger equation
    (cf. section \ref{sec:holst-prim-regime}), 2.) Keldysh
    perturbation theory based on a self consistent RPA approximation
    (cf. section~\ref{sec:keldysh}), 3.) semiclassical approach based
    on the approximate conservation of adiabatic invariants
    (cf. section \ref{sec:deepadiabaticlimit} ), 4.) an extended
    semiclassical scheme, where semiclassical methods are employed to
    propagate the full quantum distribution beyond the quantum regime
    (cf. sections~\ref{sec:twa} and \ref{sec:numerics}), and 5.)
    numerical solution of the Schr\"odinger equation for moderately
    large $N$.  }
\end{figure}

In the following, we will describe different stages of the conversion
process, where the organizing principle will be the conversion
efficiency, which in turn depends on the speed of the sweeping
process. For later reference, the different parameter regimes, along
with the theoretical methods we will use to describe them, are
summarized in Fig.~\ref{figures}.

\subsection{Alternate model representations}
\label{sec:altern-model-repr}

\begin{widetext}

\begin{figure}[h]
  \centering
  \includegraphics[width=6.3in]{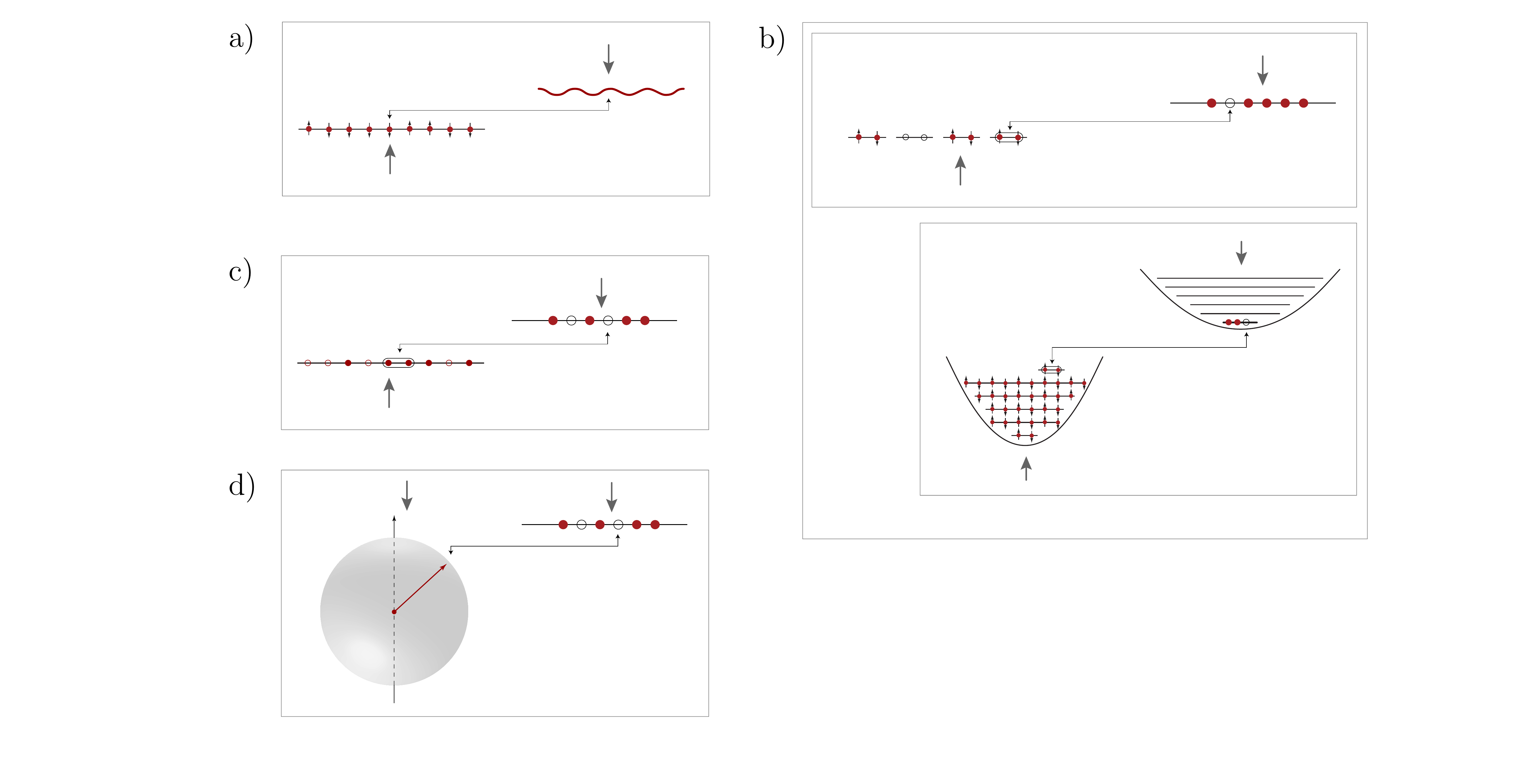}
  \caption{\label{fig:model}
   The different incarnations of the model: a) time dependent Dicke
   model: an assembly of degenerate two level systems coupled to a
   boson mode, b) flat band approximation (top) to a system of spinful
 fermions pair-binding into bosonic 'molecules', c) system of atoms
 pair-converting into molecules, d) spin subject to a magnetic field
 and coupled to a bosonic level.}
\end{figure}

\end{widetext}

The equivalence to a system of boson-coupled degenerate two level
systems -- the \textit{time dependent Dicke model} -- is best seen starting
from the latter. Consider the Hamiltonian
\begin{equation}
\label{eq:5}
\hat H = -  \lambda t ~ b^\dagger  b + \frac{\lambda t} 2 \sum_{i=1}^N
\sigma^z_i + \frac{g}{\sqrt{N}} \sum_{i=1}^N \left(  b^\dagger \,
  \sigma^-_i +  b \, \sigma^+_i \right),
\end{equation}
where $\sigma^{x,y,z}$ are Pauli matrices acting in pseudospin
space. The fact that the spin operators appear in totally symmetric
combinations, $\sum_i \sigma^{x,y,z}$ suggests introducing the total
spin operator $S^{x,y,z}\equiv {1\over 2}\sum_i \sigma^{x,y,z}$. The
operators $S^{x,y,z}$ define an ${\rm SU}(2)$-algebra. Denoting the
maximum weight space 'all fermions occupied' by $|S\rangle$, we
observe that $S^z$ has maximal eigenvalue $S= N/2$, $S^z|S\rangle = S
|S\rangle$. This shows that the operators $S^{x,y,z}$ act in a Hilbert
space of dimension $2S+1=N+1$. The global spin representation of the
Hamiltonian (\ref{eq:5}) is but the starting Hamiltonian
(\ref{eq:6}). The Hamiltonian \rfs{eq:5} is represented on
Fig.~\ref{fig:model} a).  As mentioned in the Introduction, a natural
experimental realization of \rfs{eq:5} is a cavity QED with $N$
two-level systems.
%%%%%%
%%%%%%%
%%%%%%%

The \textit{fermi-bose crossover model} describes a system of $2N$
degenerate spinful fermion states. Two fermions occupying such a state
may combine to form a boson populating one (`condensate') bosonic
state. Starting from an infinitely low initial value, the energy of
the fermions gets pushed up linearly in time
(cf. Fig.~\ref{fig:model}, b) ). This setup is described by the
Hamiltonian
\begin{eqnarray}
\label{eq:4}
\hat H = - \lambda t ~ b^\dagger
  b + \frac{\lambda t} 2 \sum_{i=1}^N \left(
    a^\dagger_{i\uparrow}  a_{i\uparrow} +
    a^\dagger_{i\downarrow}  a_{i\downarrow} \right)+ \cr +
  \frac{g}{\sqrt{N}} \sum_{i=1}^N \left(  b^\dagger \,
    a_{i\downarrow}  a_{i\uparrow} +  b \,
    a^\dagger_{i\uparrow}  a^\dagger_{i\downarrow}\right),
\end{eqnarray}
where $\lambda$ is the driving rate, $g/\sqrt N$ the coupling
amplitude of the conversion process, and $a_{i,\uparrow \downarrow}$
and $b$ are fermion and boson annihilation operators. For convenience,
we have split the time dependent energy symmetrically between bosons
and fermions. (By a gauge transformation any distribution of energies
with fixed difference $2\lambda t$ between a boson and a pair of
fermions can be realized.)

Noting that the interaction couples only two
($|\uparrow,\downarrow\rangle$ and $|0\rangle$) of four possible
($|\uparrow,\downarrow\rangle$, $|\uparrow\rangle$,
$|\downarrow\rangle$ and $|0\rangle$) occupation states of a spinful
fermion level, we may introduce a pseudospin state to discriminate
between these two states, $|\uparrow,\downarrow\rangle \leftrightarrow
\left(\begin{smallmatrix} 1\cr 0
\end{smallmatrix}\right)
$ and $|0\rangle \leftrightarrow \left(\begin{smallmatrix}
  0\cr 1
\end{smallmatrix}\right)$. The pseudospin representation of  the
Hamiltonian is but  (\ref{eq:5}).

We may think of the Hamiltonian (\ref{eq:4}) as a cartoon version of
more realistic models. For example, it can be interpreted as a
non-dispersive (flat-band) approximation to a system of free fermions
coupled to the lowest mode of an electromagnetic field in a cavity QED
setup. Alternately, it may be regarded as a cartoon version of a
BCS/BEC crossover system wherein a band of (lattice) fermions is
initially filled, the band dispersion is neglected, and only the
lowest condensate state of the bosons (`molecules') is kept
(Fig.~\ref{fig:model} b) inset). While the neglect of dispersion
effects may be acceptable~\cite{Sun:2008qm}, the assumption of full
band occupancy might not be met in experiment. (Ref.~\cite{Sun:2008qm}
argues that the full band assumption should be modeled by starting the
dynamics in incompletely `polarized' (pseudo)spin states, and we refer
the reader to this work for a discussion of this case.) For
completeness, let us mention that the problem of the dynamic BCS-BEC
crossover has been addressed in many publications and using different
approaches, beginning with \cite{Altman2005} and continuing with
Refs.~\cite{Barankov2005,Pazy2006,Pokrovsky2006,Strecher2007,Sun:2008qm}.
It is probably fair to say that a concensus on whether this problem
can indeed be reduced to the flat-band model with a single Bose mode
\rfs{eq:4} has not yet been reached in the literature.

Finally, consider a system of atoms sweeping through a Feshbach
resonance whereafter di-atomic molecules form the energetically
preferred state (cf. Ref.~\cite{Hodby:2005uf} where the conversion
processes has been realized in condensates of $^{85}\mathrm{Rb}$ or
$^{40}\mathrm{K}$ atoms.)  This situation is described by the
Hamiltonian
\begin{equation}
  \label{eq:17}
\hat H =  \lambda t d^\dagger d- {\lambda t\over 2} c^\dagger c+
{ g\over \sqrt N}(d^\dagger cc + d c^\dagger c^\dagger).
\end{equation}
This Hamiltonian  describes the \textit{conversion of atoms to
  molecules}, where the former/latter are created by the bosonic
operators $c^\dagger$/$d^\dagger$, and the symmetry $[\hat H,
d^\dagger d + 2 c^\dagger c]=0$ reflects the conservation of the total
number of particles. The time-independent version
of this model was extensively analyzed in Ref.~\cite{Radzihovsky2004}.

The (near) equivalence to the previous model
representations is seen by computing matrix elements $M_{N,n}\equiv \langle N-n+1,2n-2|\hat
H| N-n,2n\rangle$, where the state $|N-n,2n\rangle$ contains $(N-n)$
$d$-molecules and $2n$ $c$-atoms. Evaluating the matrix element, we
find
$$
M_{N,n} = { g\over \sqrt N}\sqrt{(N-n+1)n(n-1/2)}.
$$
Now, let $|N/2-n,n\rangle'$ be the state with $n$ bosons and spin
polarization $S_z = N/2-n$. Computing the matrix elements $M'_{N,n}
\equiv \langle N/2-n+1,n-1|\hat H| N/2-n,n\rangle$ of the Hamiltonian
(\ref{eq:6}), we find
\begin{equation} \label{eq:matrixelement}
M'_{N,n} = { g \over \sqrt N} \sqrt{(N-n+1)n^2}.
\end{equation}
The  matrix elements $M$ and $M'$ are identical, up to corrections of
$\mathcal{O}(n^{-1})$ which vanish in the limit of large particle
number occupancies. This consideration shows that the conversion of
atoms to molecules can be discussed within the framework of the spin
Hamiltonian above, where the number of molecules is counted in terms
of the spin quantum number.

The correspondence between models can be made perfect by considering
heteromolecular conversion processes (cf. Ref.~\cite{Papp:2006fx} for
a realization in an $^{85}\mathrm{Rb}-^{87}\mathrm{Rb}$ gas.)
Straightforward generalization of (\ref{eq:17}) obtains the  Hamiltonian
\be \label{eq:hamdifbos}
\hat H =  \lambda t d^\dagger d- {\lambda t\over 2} (c_1^\dagger
c_1+c_2^\dagger c_2)+
{ g\over \sqrt N}(d^\dagger c_1c_2 + d c_1^\dagger c_2^\dagger),
\ee
where $c_{1/2}^\dagger$ creates atoms of species $1/2$ and $d^\dagger$
the (1-2) heteromolecule. It is not difficult to verify that the
matrix elements of this Hamiltonian \textit{exactly} coincide with
those of the spin-boson Hamiltonian (\ref{eq:6}).

In this incarnation our problem has been analyzed by a number of
authors, beginning from Ref.~\cite{Yurovsky2002} (see also the more
recent Ref.~\cite{Liu2008}). It is also in this form that the problem
is the closest to existing
experiments~\cite{Hodby:2005uf,Papp:2006fx}.

Eqs.~(\ref{eq:6}), (\ref{eq:5}), (\ref{eq:4}), and (\ref{eq:17})
define four equivalent definitions of the model. For the convenience
of the reader, the correspondence between the systems involved in
these definitions is summarized in table \ref{tab:1}.

\begin{widetext}

\begin{table}
  \centering
  \begin{tabular}{|l||l|l|}\hline
    \textbf{model}&\textbf{system A}&\textbf{system B}\\\hline\hline
    spin-boson&spin $S=N/2$& boson mode\\
    time dependent Dicke model&$N$ two-level systems& boson mode\\
    fermi-bose conversion&flat band of $2N$ fermion states&boson
    mode\\
    atom-molecule conversion&di-bosonic molecular state&boson mode\\\hline
  \end{tabular}
  \caption{survey of the equivalent representations of the model}
  \label{tab:1}
\end{table}
\end{widetext}

\section{Linearized diabatic regime}
\label{sec:holst-prim-regime}

We first consider moderately fast driving rates where only a small
fraction of the particles undergoes a conversion process. In the
language of the spin model this means that at large times the spin
will still be stuck close to the north polar regions, $\lim_{t\to
  \infty} (S-S_z(t))/S \ll 1$, or $n_b/N\ll 1$. In this regime, the
curvature of the $\mathrm{SU}(2)$ spin manifold is not yet felt and
the problem can be effectively linearized.

This is best done in the Holstein-Primakoff representation of spin
operators
\begin{align}
  \label{eq:8}
  S^-&= c^\dagger (2S-c^\dagger c)^{1/2},\nonumber\\
  S^+&= (2S-c^\dagger c)^{1/2} c ,\nonumber\\
  S^z&=S-c^\dagger c,
\end{align}
where the algebra of bosonic operators, $c$ and $c^\dagger$ defines
the Holstein-Primakoff bosons. Substitution of the linearized
approximation $S^-=(2S)^{1/2}c^\dagger, S^+=(2S)^{1/2}c,
S^z=S-c^\dagger c$ into (\ref{eq:6}) generates the quadratic
Hamiltonian
\begin{equation}
  \label{eq:9}
  \hat H_{\rm HP} = -  \lambda t ~ (b^\dagger  b+c^\dagger c) +
g (b^\dagger c^\dagger + cb).
\end{equation}
To compute the number of bosons created in the linearized evolution,
we switch to the Heisenberg representation $b(t)=e^{i\hat H t} b
e^{-i\hat H t}$ and consider the expectation value
$$
n_b(t) = \langle b^\dagger(t) b(t)\rangle,
$$
taken with respect to the vacuum state, $\langle \dots \rangle \equiv
\langle 0|\dots |0\rangle$, with $b|0\rangle = c|0\rangle=0$.
(Alternatively, one can consider a model of bosonic atoms and
molecules, in the regime where most of the particles are molecules.
In the Hamiltonians \rf{eq:17} and \rfs{eq:hamdifbos} one may then
replace $d \rightarrow \sqrt{N}$, to arrive precisely at \rfs{eq:9},
an appropriate relabeling of operators understood. In
Ref.~\cite{Yurovsky2002} a model equivalent to the linearization above
was introduced, and the mean conversion rates of the linearized dynamics
were computed (see also Ref.~\cite{Kayali2003}).

The computation of $n_b=\lim_{t\to \infty}n_b(t)$ is reviewed in Appendix
\ref{sec:derivation-eqs.-xx}, with the principal result
\begin{equation}
  \label{eq:15}
  n_b = x-1.
\end{equation}
Eq.~(\ref{eq:15}) states that the number of converted particles grows
linearly in the LZ parameter. Unlike with the standard (two-level) LZ
problem, the many particle system remains parametrically detached from
adiabacity, even for large values of the LZ parameter. However, the
linearization underlying the derivation limits the validity of the
result to $n_b \ll N$.

The techniques used in Appendix \ref{sec:derivation-eqs.-xx} actually
yield the entire (quantum) distribution of the number of produced
particles,
\begin{equation}
  \label{eq:19}
  P(n) = {1\over x} (1-x^{-1})^n\stackrel{x\gg 1}\simeq {e^{-n/x}\over x}.
\end{equation}
This distribution is very broad. Its width $(\langle n^2\rangle -
\langle n\rangle^2)^{1/2} \simeq x$ is of the same order as its
mean. This is remarkable inasmuch as our system is not subject to any
sources of fluctuations, other than its intrinsic quantum
fluctuations. We are met with the unusual situation that quantum
fluctuations are of the same order as mean values, in spite of
the 'semiclassical' largeness of the latter,  $\langle n \rangle
\simeq x\gg1$. Similar phenomena have been discussed in the context of
dynamical sweeps through quantum phase transition points,
cf. e.g. Ref.~\cite{schutz2005}.

\section{The Kinetic Equation}
\label{sec:keldysh}
\subsection{Keldysh formalism}

An alternative way of analyzing the problem starts from the
boson-fermion representation (\ref{eq:4}) and analyzes the
corresponding kinetic equation -- theory strand b) in
Fig.~\ref{fig:model}. To formulate this approach, we employ the
Keldysh formalism and introduce the appropriate Keldysh Greens
functions (cf. Ref.~\cite{Kamenev2004} for a review of the formalism.)
Readers not interested in the formalism are invited to proceed
directly to the discussion of the main result of this section, the
rate equation (\ref{eq:34}).

Within the Keldysh formalism, the fermion subsystem is described by
the matrix Greens function
\begin{align*}
&G_{\sigma \alpha,\sigma' \alpha'} (t,t') = -i \left< a^{\vphantom{\dagger}}_{\sigma
    \alpha} (t) a^\dagger_{\sigma'\alpha'} (t') \right> =\\
&\qquad=
 \left( \begin{matrix}  G^R  & G^K \cr 0 & G^A  \end{matrix} \right)_{\sigma\alpha,\sigma'\alpha'}\hspace{-.5cm}(t,t'),
\end{align*}
where $\alpha,\alpha'=1,2$ index the two-component Keldysh space (to
which the matrix structure in the equation refers), and
$\sigma=(i,\uparrow\downarrow)$ is a container index comprising the
fermion site and spin index. $G^R$, $G^A$, and $G^K$ stand,
respectively, for the advanced, retarded and Keldysh Greens
function. The latter is defined by
$$
G^K = G^R \circ f - f \circ G^A,
$$
where the symbol $'\circ'$ defines a temporal convolution,
$$
(A\circ B)(t,t')\equiv \int dt^{\prime\prime}\, A(t,t^{\prime
  \prime})B(t^{\prime\prime},t')
$$
and $f\equiv 1-2n_f$ defines the distribution of
the fermions. In what follows, it will be convenient to represent the
Greens functions of the theory in the Wigner representation,
\begin{equation}
G^K(\tau,\omega)=\int dt~G^K\left(\tau+\frac t 2,
\tau- \frac t 2\right) e^{i \omega t},
\end{equation}
and similarly for all other Green functions.  In the absence of
interactions, $g=0$, a straightforward calculation obtains the Wigner
representation of the free Greens functions as
\begin{eqnarray} \label{green1}
G_{0,\sigma\sigma'}^R(\tau,\omega) & =&
{\delta_{\sigma\sigma'}\over \omega^+-\lambda t/2},\\
G_{0,\sigma\sigma'}^A(\tau,\omega) & =&  { \delta_{\sigma\sigma'}\over
  \omega^- - \lambda t/2},\cr
G_{0,\sigma\sigma'}^K(\tau,\omega) & =&  -2\pi
\delta_{\sigma\sigma'}f_0(\tau,\lambda t/2) \delta(\omega-\lambda t/2),\nonumber
\end{eqnarray}
where $f_0(\tau,\lambda t/2) \equiv f_0(\tau)$ describes the
distribution of the full fermion band, $f_0(\tau) = 1-2n_0(\tau) =
-1$.  In a similar manner the bosons are described by
\begin{equation}
D(t,t') = -i \langle b^{\vphantom{\dagger}}_\alpha(t)  b^\dagger_{\alpha'}(t') \rangle=
\left( \begin{matrix}  D^K & D^R \cr D^A & 0  \end{matrix}
\right)_{\alpha\alpha'}\hspace{-.5cm}(t,t'),
\end{equation}
where
$$
D^K = D^R \circ F - F\circ D^A,
$$
and $F=1+2n_b$, $n_b$ describing the boson distribution.
% At large
% negative times, $n_b(t,t')\stackrel{t,t'\to -\infty}{\longrightarrow}
% \delta(t-t') n_b^0$, $n_b^0=0$, no bosons are present.
The Wigner representation of the
non-interacting boson Greens functions is given by
\begin{eqnarray} \label{green2}
D_0^R(\tau,\omega) & =&  {1\over \omega^+ + \lambda t},\\
D_0^A(\tau,\omega) & =&  {1\over \omega^- + \lambda t},\cr
D_0^K(\tau,\omega) & =&  -2\pi F_0(\tau,-\lambda t)\,
\delta(\omega+\lambda t),\nonumber
\end{eqnarray}
where $F_0(\tau,-\lambda \tau)\equiv F_0(\tau)$ relates to the vanishing boson number
of the non-interacting problem as $F_0(\tau) = 1+2n_0(\tau)=1$.
%where $1+2n_b^0=1$.

% Together with the Greens functions, we introduce the bosonic and
% fermionic distribution functions $F(t,t')$ and $f(t,t')$ defined by
% \begin{equation}
% D^K(t,t') = \int dt'' \left[ D^R(t,t'') F(t'',t') - F(t,t'')
%   D^A(t'',t') \right],
% \end{equation}
% and similarly
% \begin{equation}
% G^K(t,t') = \int dt''  \left[ G^R(t,t'') f(t'',t') - f(t,t'')
%   G^A(t'',t') \right].
% \end{equation}
% In what follows, it will be advantageous to recast all the Greens
% function in the form of their Wigner transform, defined by
% \begin{equation}
% G^K(\tau,\omega)=\int dt~G^K\left(\tau+\frac t 2,
% \tau- \frac t 2\right) e^{i \omega t}
% \end{equation}
% and by identical expression for all other bosonic and fermionic Greens
% and distribution functions. Comparison with \eqref{green1} shows that
% $$
% G^K(\tau,\omega) = 2\pi i  f_0(\tau,\lambda t/2) \delta(\omega-\lambda t/2),
% $$

In order to calculate the number of produced bosons at the end of the
Landau-Zener process, we need to calculate the exact bosonic Keldysh
Greens function, in terms of which
\begin{align} \label{eq:kr}
n_b &= \lim_{t \rightarrow \infty} n_b(t) = \lim_{t \rightarrow \infty}
\left( iG^K(t,t)-1 \right)/2\simeq\cr
&\simeq{1\over 2}\left(\lim_{\tau \rightarrow \infty}  \int {d\omega\over 2\pi} F(\tau,\omega)
A_b(\tau,\omega) -1\right),
\end{align}
where $A_b = -2 {\rm Im}\,D^R$ is the spectral function of the bosons.
For large positive times, the Greens functions
become effectively uncoupled, $D^R(\omega)$ asymptotes to the free form
\eqref{green2} and $A_b(\omega) \simeq 2\pi \delta(\omega +\lambda t)$. This means that the number of bosons at the end of the
process is given by
\begin{equation}
  \label{eq:distden}
n_b = \lim_{\tau\to \infty} {F(\tau,-\lambda \tau)-1\over 2}.
\end{equation}

\subsection{Perturbation theory}

First, we can try to calculate the bosonic Greens function
perturbatively, in powers of $g$ (which means, in practice, in powers
of the dimensionless parameter $g^2/\lambda$). This expansion is valid
at fast rate $\lambda$, $g^2/\lambda \ll 1$.
\begin{figure}[h]
  \centering
  \includegraphics[height=2.5in]{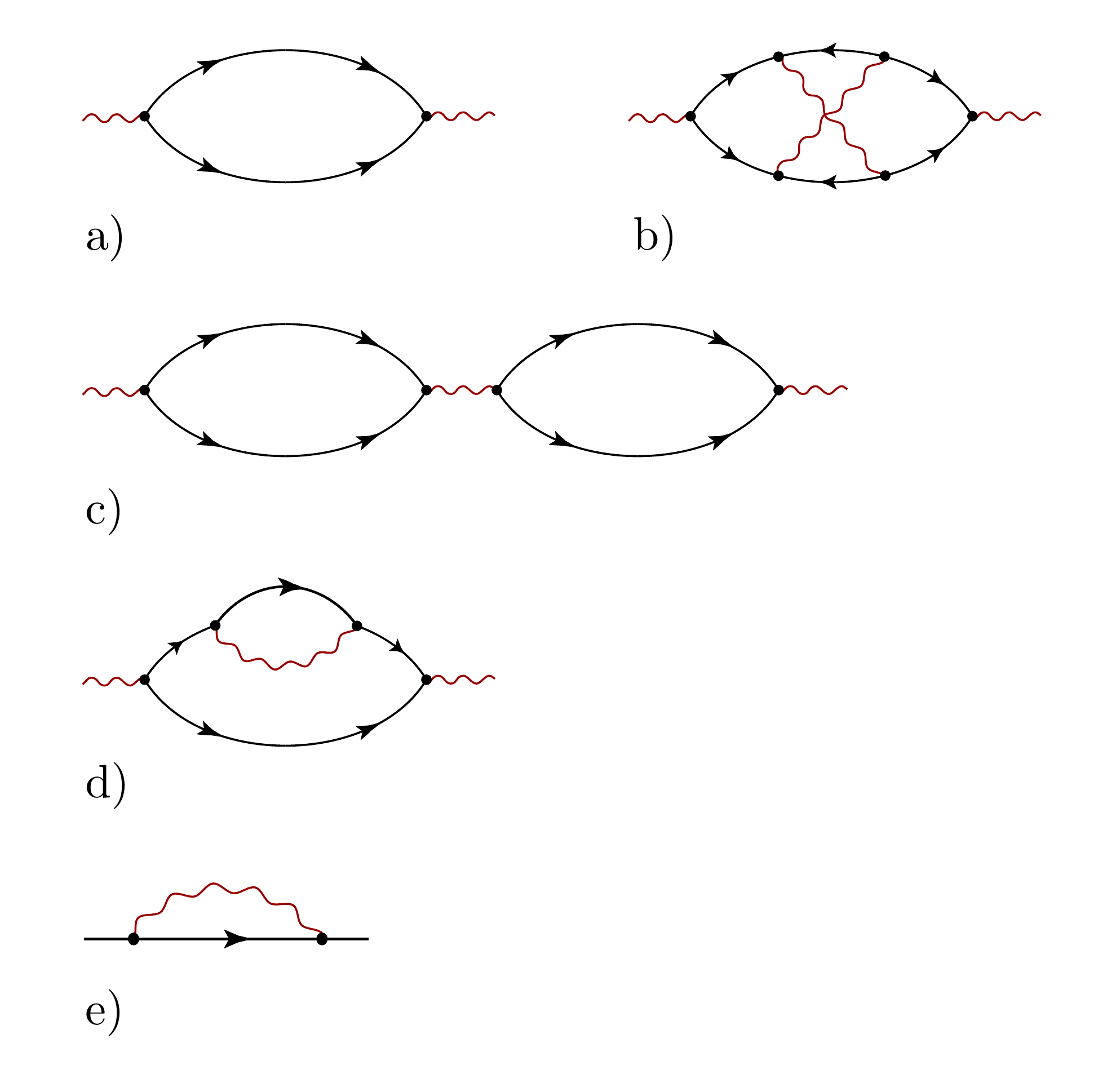}
  \caption{\label{self_energy_lowest}
The lowest order corrections to the bosonic propagator.}
\end{figure}
The lowest order correction to the bosonic propagator is depicted in
Fig.~\ref{self_energy_lowest}, diagram a). The calculation of this
diagram, although cumbersome, proceeds in a straightforward fashion,
giving the result
\begin{equation}
n_b  \approx \frac{\pi g^2}{\lambda}.
\end{equation}
Each vertex of the diagram carries a factor of $g/\sqrt{N}$. However,
there are $N$ fermions propagating around the loop, canceling the
factor $1/N$ coming from the verices. This results in the cancellation
of factors of $N$ in the answer.

In the next order of perturbation theory two more diagrams contribute
to the bosonic propagator. These are shown on
Fig.~\ref{self_energy_lowest}, diagrams c) and d). This leads to the
second order answer
\begin{equation}\label{eq:taylor}
n_b  \approx \frac{\pi g^2}{\lambda} + \left(\frac 1 2 - \frac 1 N  \right) \left ( \frac{\pi g^2}{\lambda} \right)^2.
\end{equation}
Note that at $N=1$, this indeed coincides with the Taylor expansion of
the exact Landau-Zener formula
\begin{equation} n_b= 1 - e^{-\frac{\pi g^2}{\lambda}}.\end{equation}
Notice also that at $N>1$, the number of produced particles is bigger
than at $N=1$. This is natural, as with larger $N$ one can produce
more bosons.

Finally, we note that the calculation leading to Eq.~(\ref{eq:taylor})
closely parallels that employed in Ref.~\cite{Pokrovsky2006} to calculate the molecule
production in a Feshbach resonance experiment perturbatively.

\subsection{Self consistent RPA and kinetic equation}
\label{sec:self-consisitent-rpa}

Our aim now is to calculate the number of produced bosons in
the large $N$ limit, with all other parameters including $g$ and
$\lambda$ kept fixed. Our starting point is the system of kinetic equations (the
Wigner transforms of the Dyson equations, cf. Ref.\cite{Kamenev2004})
for the boson and fermion distribution function,
\begin{align}
\label{eq:20}
\left(  \partial_\tau  - \lambda \partial_\omega \right) F(\tau,
\omega) &= i\left(\Sigma^K -  \left( \Sigma^R \circ F - F \circ \Sigma^A
\right)\right)(\tau,\omega),\\
\left(  \partial_\tau +  {\lambda\over 2} \partial_\epsilon \right) f(\tau,
\epsilon) &= i\left(\sigma^K -  \left( \sigma^R \circ f - f \circ \sigma^A
\right)\right)(\tau,\epsilon),
\end{align}
The first of these equations is obtained by evaluating the Wigner
transform of the general kinetic equation of a boson
system~\cite{Kamenev2004} $ (- i \partial_\tau + [\hat H,\;]) F =
\Sigma^K - \left( \Sigma^R \circ F - F \circ \Sigma^A \right)$ for the
Hamiltonian $\hat H = - \lambda t$. The second equation is its
fermionic analog.

In the 'collision integrals' of \eqref{eq:20}, $\Sigma^{K,R,A}$
($\sigma^{K,R,A}$) are the components of the bosonic/fermionic self
energies. It is not difficult to see that the dominant contributions
to these self energies are given by the diagrams
Fig.~\ref{self_energy_lowest} a) and e), external legs
truncated. Unlike with the lowest order perturbation theory discussed
above, the fermionic and bosonic propagators appearing in these
diagrams are to be understood as dressed propagators, containing self
energy insertions by themselves. (Technically speaking, this means
that we will evaluate the self energies in a self consistent RPA
approximation.) Self energy diagrams with crossing interaction lines,
such as Fig.~\ref{self_energy_lowest} b),  carry factors of
$N^{-1}$ relative to the self consistent RPA diagrams. (See, however,
the discussion in the end of this section.)

The derivation of the self energies for the boson-fermion interaction
follows  standard procedures~\cite{Kamenev2004} and we will not
repeat it here. As a result we obtain the equations
\begin{widetext}
\begin{eqnarray} \label{eq:longkin}
\left( \d_\tau - \lambda \d_\omega \right) F(\omega) &=&  {g^2\over 2} \int
{d\epsilon\over 2\pi }~ A\left(\omega-\epsilon-\frac{\lambda \tau}{2}\right)
A\left(\epsilon-\frac{\lambda \tau}{2} \right) \left[ 1+f(\omega-\epsilon) f(\epsilon) -
\left(f(\epsilon) + f(\omega-\epsilon) \right) F(\omega) \right] , \cr
\left( \d_\tau + \frac{\lambda}{2} \d_\epsilon \right) f(\epsilon) &=& {g^2\over
  2N}\int
{d\omega\over 2\pi}~ A \left(\omega-\epsilon-\frac{\lambda \tau}{2}\right)
A_b\left(\omega+ {\lambda \tau} \right) \left[ 1+f(\omega-\epsilon) f(\epsilon) -
\left(f(\epsilon) + f(\omega-\epsilon) \right) F(\omega) \right],
\end{eqnarray}
\end{widetext}
where we suppressed the explicit $\tau$-dependencies for notational
clarity and $A=-2 {\rm Im}\,G^R$ and $A_b= -2 {\rm Im}D^R$ are the
spectral functions of the fermions and the boson, respectively. Notice
the factor of $N^{-1}$ multiplying the collision integral of the
fermions. The absence of this factor in the boson collision integral
signals that the boson interacts with all $N$ fermions
simultaneously. In contrast, each fermion interacts only with a single
boson, which means that the fermionic self energies carry the
uncompensated squared coupling constant $g^2/N$.

Eq.~\eqref{eq:longkin} defines a particle number conserving
system. Consider the quantity $N(\tau) \equiv n_b(\tau) + N n_f(\tau)$. Using
Eq.~\eqref{eq:kr}, and its fermionic analog $n_f(\tau) = -{1\over
  2}\left(\int {d \epsilon\over 2\pi} f(\tau,\epsilon) A(\tau,\epsilon) -1
\right)$, we obtain that the number of particles varies in self
consistent RPA as
$$
{d N\over d\tau} = {1\over 2}d_\tau\Big(  \int {d\omega\over 2\pi} F(\tau,\omega)
A_b(\tau,\omega) - N\int {d \epsilon\over 2\pi} f(\tau,\epsilon)
A(\tau,\epsilon)\Big).
$$
Integrating the first (second) equation \eqref{eq:longkin} over $\int
{d\omega\over 2\pi} \,A_b(\tau,\omega)$ $\left(\int {d\epsilon\over
  2\pi} \, A(\tau,\epsilon)\right)$ and using that
$\lambda \partial_\omega A_b = \partial_\tau A_b$, we find that the
two integrals cancel out, and $d_\tau N=0$.

This is about as far as we will get without further
approximations. Below we will argue that the validity of the RPA
approximation is limited by the condition $n_b \ll N$. In the limit
$n_b/N \to 0$, the self energies $\sigma^{R/A}$ broadening the energy
dependence of the spectral functions become vanishingly small. (The
detailed dependence of the broadening depends on the value of time,
$\tau$.) Relying on this limiting behaviour, we will approximate the
spectral functions by $A(\epsilon) \simeq 2\pi \delta(\epsilon)$ by
$\delta$-functions. This is a bold approximation inasmuch as the pairs
of $\delta$-functions in \eqref{eq:longkin} enforce the resonance
condition $\tau=0$. However, it is this 'resonant' time window, $\tau
\simeq0$ where the distribution functions entering the integral kernel
are expected to vary strongest. This means that the value of the
collision integral may well be sensitive to the detailed temporal
profile of the distribution functions around $\tau=0$. Ultimately, we
will need to compare to the result of other methods to verify the
validity of the $\delta$-function approximation.

We substitute the approximations $A_b(\tau,\omega) \simeq 2\pi
\delta(\omega + \lambda \tau)$ and $A(\tau,\epsilon) \simeq 2\pi
\delta(\epsilon- \lambda \tau/2)$ into \eqref{eq:longkin}, use
$n_b(\tau) = {1\over 2} \left(\int {d \omega\over 2\pi} F(\tau,\omega)
  A_b(\tau,\omega) -1 \right)\simeq (F(\tau,-\lambda \tau)-1)/2$, and
differentiate w.r.t. time to obtain
\begin{equation} \label{eq:34}
\d_\tau n_b = - \d_\tau n_f = \frac{\pi g^2}{\lambda} \delta(  \tau) \left[  n_f^2 + n_b (2 n_f-1) \right].
\end{equation}
The meaning of this equation becomes transparent upon rewriting the
combination of distribution functions on its r.h.s as $(1+n_b) n_f^2 -
n_b (1-n_f)^2$. This factor weighs the probability that two fermions
convert into bosons and back. To solve the equation, we take advantage
of the conservation law
\begin{equation}
n_b + N n_f = N.
\end{equation}
It is then straightforward to obtain the solution
\begin{widetext}
\begin{equation}
  n_b (\tau)= \frac{ 2N \left( e^{\sqrt{\frac{N+4}N} \frac{\pi
          g^2}{\lambda} \theta(\tau)}-1 \right) }{
    e^{\sqrt{\frac{N+4}N} \frac{\pi g^2}{\lambda}\theta(\tau)} \left(
      - N + \sqrt{N (N+4)} +2 \right) +N +
    \sqrt{N (N+4)} -2 }.
\end{equation}
\end{widetext}
The singular behaviour of $n_b$ at $\tau=0$ is a remnant of the above
$\delta$-function approximation. Incidentally, we note that the Taylor series
expansion of $n_b$ with respect to $g^2$ gives
\begin{eqnarray}
n_b =\lim_{\tau\to \infty}n_b(\tau)&=&  \frac{\pi g^2}{\lambda} + \left( \frac 1 2 - \frac 1 N \right) \left( \frac{\pi g^2}{\lambda} \right)^2 + \cr  && + \frac{6-8 N + N^2}{6 N^2} \left(
\frac{ \pi g^2}{\lambda} \right)^3 + \dots
\end{eqnarray}
The first two terms indeed match the direct perturbative expansion
given by \rfs{eq:taylor}, confirming that this technique does take
into account all the diagrams up to the second order. (In the context
of perturbation theory, the $\delta$-function approximation of the
spectral functions amounts to the lowest order in $g$ Born
approximation to the self energy operator.)

On the other hand, taking the limit $N \rightarrow \infty$ gives
\begin{equation} \label{eq:kinan}
n_b = \frac{N \left( e^{\frac{\pi g^2}{\lambda} }-1 \right)}{2 e^{\frac{\pi g^2}{\lambda} }+ N }.
\end{equation}
This is the main result of the  kinetic equation approach.
For  $e^{\pi g^2/\lambda} \ll N$ we find
\begin{equation} \label{eq:kinanHP}
n_b = e^{\frac{\pi g^2}{\lambda} }-1.
\end{equation}
This matches the large $N$ limit given by \rfs{eq:15}.

The kinetic equation produces the boson number \rf{eq:kinan} which is
monotonously increasing with decreasing $\lambda$, until it reaches
its limiting small $\lambda$ value of $N/2$. The reason for this
behaviour is that at large $N$, the r.h.s. of the kinetic equation
approaches zero at $n_f=1/2$. In other words, the rate of fermions
converting into bosons is matched by the rate of bosons converting
into fermions. However, we need to remember that at $n_b=N/2$ we are
well outside the regime of validity of the approximated kinetic
equation. In fact, we are outside the limit of the RPA as such. This
is because at $n_b\simeq N$ diagrams that are nominally small in
$N^{-1}$ become sizeable, on account of the large value of the
distribution function $n_b \simeq N$. For example, it is
straightforward to verify that its crossing interaction lines make
diagram Fig.~\ref{self_energy_lowest} b) small in $N^{-1}$, but this
smallness gets counteracted by $n_b$-dependent propagators in the
center regions of the diagram. The bottom line is that the RPA is
controlled only in the limit $n_b/N\to 0$. Our comparison with numerics
below will show that it works remarkably well beyond its nominal
limits, i.e. the profile \eqref{eq:kinan} represents a good
approximation even at $n_b/N =\mathcal{O}(1)$. However, in order to
advance into the adiabatic regime, $n_b \to N$, we need to employ
different methods.

\section{Semiclassical  analysis. }
\label{sec:twa}

Our  goal now is to understand the behavior of $n_b$ for arbitrary driving
rates $\lambda$. We
will use that
for  large $N\gg 1$ our problem approaches a well defined classical limit where
both spin and the bosonic fields can be treated as classical objects. The
classical equations of motion, obtained by replacing operators in the
Hamiltonian~(\ref{eq:6}) by $c$-numbers and commutators by Poisson
brackets, read
\begin{align}
  &i\dot b=-\lambda t b+{g\over\sqrt N}S^-, \cr
&\dot S_z=-i{g\over\sqrt{N}}(bS^+-b^\ast S^-),\cr
&\dot S_x=-i{g\over\sqrt{N}}(b^\ast-b)S_z-\lambda t S_y\cr
&\dot S_y=-{g\over\sqrt{N}}(b+b^\ast)S_z+\lambda t S_x.
\label{cl_eq4}
\end{align}
Within the strictly classical setting, these equations have to be
solved for the initial conditions: $b(t_0)=0$, $S_z(t_0)=N/2$,
$S_x(t_0)=S_y(t_0)=0$, where $t_0\to -\infty$ is the initial
time. It is straightforward to verify that the solution corresponding
to these initial conditions reads $(b(t),S_z(t)) = (b(t_0),S_z(t_0)) =
(0,N_0)$, i.e. the initial configuration of zero bosons represents a
classically stationary (if unstable) solution.

The prediction that no bosons will be generated, is clearly
wrong. However, the purely classical treatment has yet another, and
related, drawback: for given initial conditions the uniqueness of the
classical solution does not permit the buildup of fluctuations. This
is in contradiction with our earlier findings that the distribution of
$n_b$ is wide at least at fast and intermediate rates.

 It is thus obvious that an meaningful description of the
classical limit must account for the presence of quantum
fluctuations. These fluctuations will lead to an initial
destabiliziation of the configuration $(b(t),S_z(t)) = (0,N_0)$. The
nonlinearity of the classical Hamiltonian, then amplifies the
fluctuations and drives the system away from its initial state. The
expansion of dynamics in quantum fluctuations for bosons was analyzed
in detail in Ref.~\cite{ap_twa}. Using a path integral
construction similar to the Keldysh technique it was shown that to
leading order in quantum fluctuations the classical (Gross-Pitaevskii)
equations of motion do not change. However, the initial conditions become
randomly distributed with the weight given by the Wigner transform of
the density matrix. This so called
Truncated Wigner Approximation (TWA)  originated in quantum
optics~\cite{walls-milburn, gardiner-zoller} and has later been
adopted to cold atom systems~\cite{walls}. By by now the
TWA has become a standard tool in describing the  dynamics of cold
atoms in  regimes where classical Gross-Pitaevskii equations do not
suffice (see Ref.~\cite{blakie} for  review).

% The further corrections to TWA appear in the form of quantum jumps on
% classical trajectories~\cite{ap_twa}.

The TWA is most conveniently  adopted to our  model by using
the Schwinger boson representation of spins. The small parameter of this
expansion is $1/S$ where $S\to\infty$ describes the
classical spin. In Appendix~\ref{app:twa} we give some details of this
derivation. The net result is that the classical equations of
motion~(\ref{cl_eq4}) need to be supplied by stochastic
initial conditions distributed according to the Wigner function. For
large spin (pointing along $z$-direction) and vacuum state of bosons,
representing the initial conditions in our problem,
this Wigner function is approximately given by
\begin{equation}
  W(n_b, S_z,
  S_\perp)\approx {2\over\pi S}\exp[-2n_0]\exp\left[-{S_\perp^2\over
      S}\right]\delta(S_z-S),
\label{wig_func}
\end{equation}
where $S_{\perp}=\sqrt{S_x^2+S_y^2}$ and $n_0$ is the initial boson
number. The fact that $S_{\perp}$ and $n_0$ are not exactly equal to
zero is the result of vacuum quantum fluctuations. In this work we are
primarily interested in the number of bosons created during the
dynamical process, and its fluctuations. The connection between
moments $n_0^l,l=1,2,\dots$ and the variables $b$ and $b^\dagger$ is
established by the Wigner transform identities summarized in
Appendix~\ref{app:twa}.  For instance,
\begin{equation}
  n_0\to b^\ast b-1/2,\quad
n_0^2\to (b^\ast b)^2-b^\ast b.
\label{weyl}
\end{equation}
(The correction term $b^\ast b$ appearing in the expression for the
second moment ensures that zero point fluctuations do not affect the
expectation value of $n_0$ and $n_0^2$, which should be zero in the
vacuum state. In the classical limit $n_0\gg 1$, these terms become
inessential.)

By way of an illustration, let us outline how the TWA reproduces the
results of the quantum calculation in the linearizable regime. For
quadratic Hamiltonians the TWA is exact~\cite{walls-milburn, ap_twa}
and we must be able to reproduce all the results of
Sec.~\ref{sec:holst-prim-regime}. Upon leaving the linear regime, the
number of bosons $n(t) = \left< b^\dagger(t) b(t) \right>$ has become
large and semiclassical methods are expected to work with high
accuracy. In other words, the exactness of the TWA to the linear
regime entails its applicability to the full parameter space of our
problem.

In the linear regime, we may assume approximate constancy of $S_z$ :
$S_z\approx S=N/2$.  Eq.~(\ref{cl_eq4}) then reduces to
\begin{align}
  \label{linear_semicl}
  i\dot b&=-\lambda t b +g s^-\cr
i\dot s^-&=-g b+\lambda t s^-,
\end{align}
where we introduced $s^-=S^-/\sqrt{N}$. These equations are identical
to an exact reformulation of the Schr\"odinger equation
(cf. Eq.~(\ref{eq:end}) in Appendix~\ref{sec:derivation-eqs.-xx}), a
consequence of the linearity of the problem. In Appendix
\ref{sec:twa-vs-quantum} we show that the solution of these equations,
with account for quantum fluctuations in the initial data, indeed
reproduces the results of the full quantum calculation. In particular
we show there that $n_b=x-1$ and $\langle n_b^2\rangle=2x^2-3x+1$
(here as before, we reserve the notation $n_b$ for the number of
bosons at the end of the process, as in \rfs{eq:7}). Both results
perfectly agree with the exact distribution~(\ref{eq:19}).

At slow values of driving, $x\sim N$ the nonlinear equations
(\ref{cl_eq4}) no longer afford an exact solution.  In the next
section we will discuss an approximate solution scheme, based on
the method of adiabatic invariants. In Sec.~\ref{sec:numerics}
we will analyze in detail $n_b(\lambda)$ and its distribution in
various regimes using both exact and TWA simulations.

\section{Deep Adiabatic Limit}
\label{sec:deepadiabaticlimit}
\subsection{Adiabatic Hamiltonian}
In this section, we will apply the concept of classical adiabatic
invariance to the adiabatic limit of the driving process. Considering
a fixed initial value $n_0 \sim 1$ (consistent with \rfs{wig_func}),
we will be sloppy about initial conditions. This is good enough to
obtain reliable results for the mean value of produced bosons,
$\langle n_b\rangle$, but won't suffice to obtain the statistics of
the driving process. The latter will be explored in section
\ref{sec:numerics} below within the more sophisticated TWA scheme.

The concept of the adiabatic evolution of classical invariants
\cite{LL1} was previously applied to the semiclassical limit of a
non-linear Schr\"odinger equation ~\cite{Niu2002} and later
\cite{Pazy2006} to the classical equations of our model. However, (for
all but very small, compared with $N$, initial occupancy $n_0 \sim 1$)
the latter reference predicts a power law $\sim \lambda^{1/3}$ for the
conversion rate which we can exclude on the basis of direct numerics,
TWA, and our analytical calculation below. We therefore believe that
the actual technical calculation of Ref.~\cite{Pazy2006} is
flawed. However, the conceptual basis for the emergence of power laws
(as opposed to exponentials such as in the original Landau Zener
problem) remains valid: in driven classical dynamical systems, certain
invariants of the autonomous limit of the evolution (viz. the action
picked up upon traversal of periodic trajectories) become weakly time
dependent. While in most regions of phase space this time dependence
is very (exponentially in the driving rate) weak, it can become strong
(algebraic) in the vicinity of singular points. Such points reflect
the presence of nonlinearities in the Hamiltonian, which in our
problem are due to interactions. When the driving parameter sweeps
across a singular point, the topology of trajectories in phase space
(and thence the action picked up along these trajectories) may change
profoundly. It is the dynamics of these processes which we will
investigate in the following.

Specifically, we will consider the system in the following limit,
\begin{equation}
   \label{eq:limit}
N \gg 1, \ \frac{\lambda}{g^2} \ll 1, \  \frac 1 {N^\alpha} \ll  \frac{\lambda}{g^2} \log N \ll 1,
\end{equation}
where $\alpha>0$ is an arbitrary positive number.

We consider the Hamiltonian (\ref{eq:6}) in the limit of $c$-number
valued operators. Using a number-phase decomposition of the boson
field, $b\to\sqrt{n}\mathrm e^{i\varphi}$, and a  polar representation
of spin variables, $S_z=S\cos\theta$, $S_x=S\sin\theta \cos\xi$,
$S_y=S\sin\theta\sin\xi$, it assumes the form
$$
H(n,\varphi,\theta,\xi)=-\lambda t n + \lambda t \frac{N}{2} \cos \,
\theta + g \sqrt{N n} \sin\left(\theta \right) \cos(\xi-\varphi).
$$
The two angles $\xi$ and $\varphi$ can be combined into one angle
$\phi=\xi-\varphi+\pi$ (where $\pi$ is added for later
convenience). The angle $\theta$ and the boson number $n$ are not
independent, but are related via the conservation law
\begin{equation} \label{eq:42}
n= \frac{N}{2} \left(1 - \cos \, \theta \right).
\end{equation}
This allows to trade $\theta$ for $n$. It is then convenient to
rescale \be \label{eq:rescale} n \rightarrow N n, \ee so that the new
$n$ varies from $0$ to $1$, with the last value taken when all
particles are bosons. Similarly, the rescaled initial condition reads
$$n_0 \sim \frac{1}{N}.
$$
It is also advantageous to replace
\be \label{eq:rescaleH} H \rightarrow N H. \ee Throughout this section, we will work with the rescaled
variables as described here, restoring the original variables at the end. 

Finally, without any
loss of generality we can set $g=1$ since it can be scaled out by the
appropriate rescaling of $t$.  We arrive at the Hamiltonian
\begin{equation} \label{eq:43}
H =-\gamma n - 2 n \sqrt{1-n} \cos(\phi), \ \gamma=2\lambda t.
\end{equation}
Here $\phi$ and $n$ play the role of the coordinate and momentum,
canonically conjugate to each other.  The equations of motion of these
variables read as
\begin{eqnarray} \label{eq:eqmh}
  d_t n & = -\partial_\phi H,\cr
  d_t \phi&=\partial_n H.
\end{eqnarray}
These equations of motion need to be supplied with initial conditions.
In this section, where  the emphasis is on the calculation of the 'typcial'
number of produced bosons, we will supply the dynamical system
(\ref{eq:eqmh}) with a fixed value of the initial $n$. We take
this to be the typical value $n_0 \sim 1/N$ of the (rescaled) distribution
(\ref{wig_func}). (As we will see below in the slow limit
$n(t\to\infty)$ has only logarithmic sensitivity to $n_0$ and precise
form of the initial value is not important, as long as we are
interested in estimating mean conversion rates. For the same reason the
initial fluctuations in the angle $\theta$, which we ignored, are also
unimportant.)

% Note that the
% extra factor of $N$ in the exponent compared to Eq.~(\ref{wig_func})
% appears because of the rescaling $n\to nN$.  In this form it is clear
% that the initial quantum fluctuations of $n$ indeed vanish in the
% classical limit $N\to\infty$. For our analytic analysis, it even
% suffices to substitute the distribution of $n(t_0)$ (which we denote
% $n_0$ throughout this section) by its typical value \be n_0 \sim
% \frac{1}{N}.  \ee

% The Hamiltonian \rfs{eq:43} has been studied in previous
% work~\cite{Liu2008}. In that reference, the atoms/molecules conversion
% problem formalized by Eqs.~\rf{eq:17} and \rf{eq:hamdifbos} was
% reduced to \rfs{eq:43} just as we did above. It was subsequently
% analyzed by means of an uncontrolled approximation, the path we do not
% take here. To analyze this Hamiltonian we instead take advantage of
% the small parameter defined in \rfs{eq:limit} and employ a procedure
% similar to the one developed in Refs.~\cite{Niu2000,Niu2002} for a
% different problem.

\begin{figure}[hbt]
\includegraphics[height=2.3in]{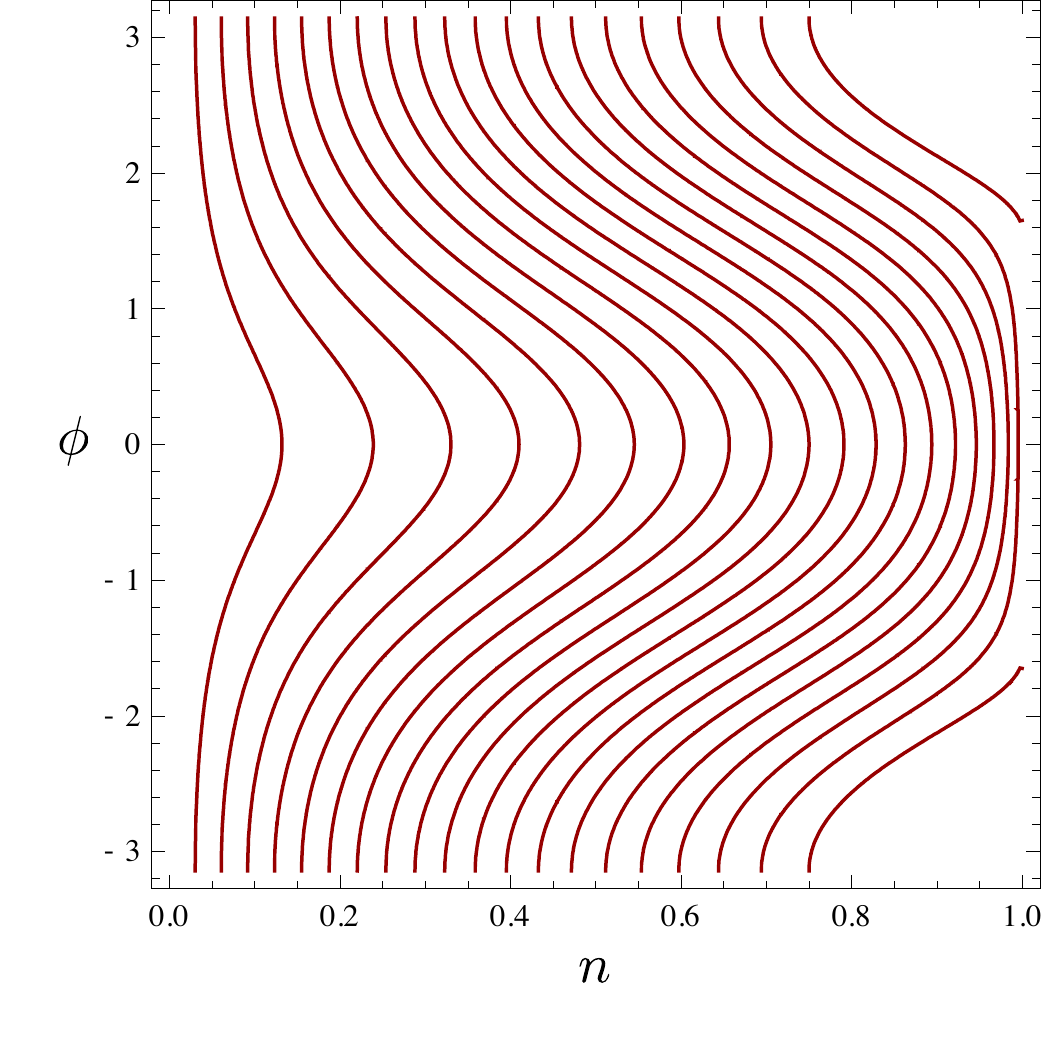}
 \caption{  \label{graph3-1} Trajectories of $H$ at $\gamma=-3$. Here the horizontal axis is $n$, and the vertical axis is $\phi$. }
\end{figure}

\begin{figure}[hbt]
\includegraphics[height=2.3in]{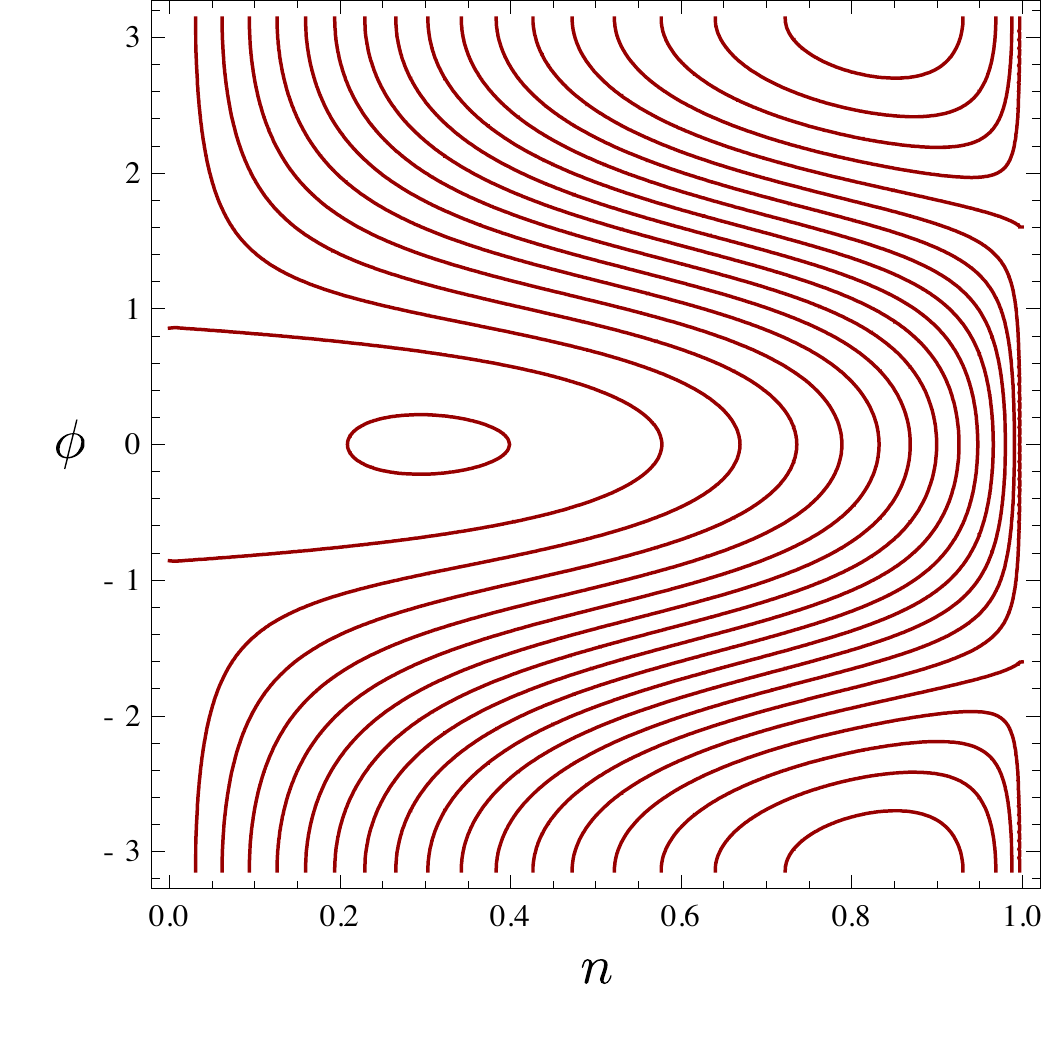}
 \caption{ \label{graph3-2} Trajectories of $H$ at $\gamma=-1.3$}
\end{figure}

\begin{figure}[hbt]
\includegraphics[height=2.3in]{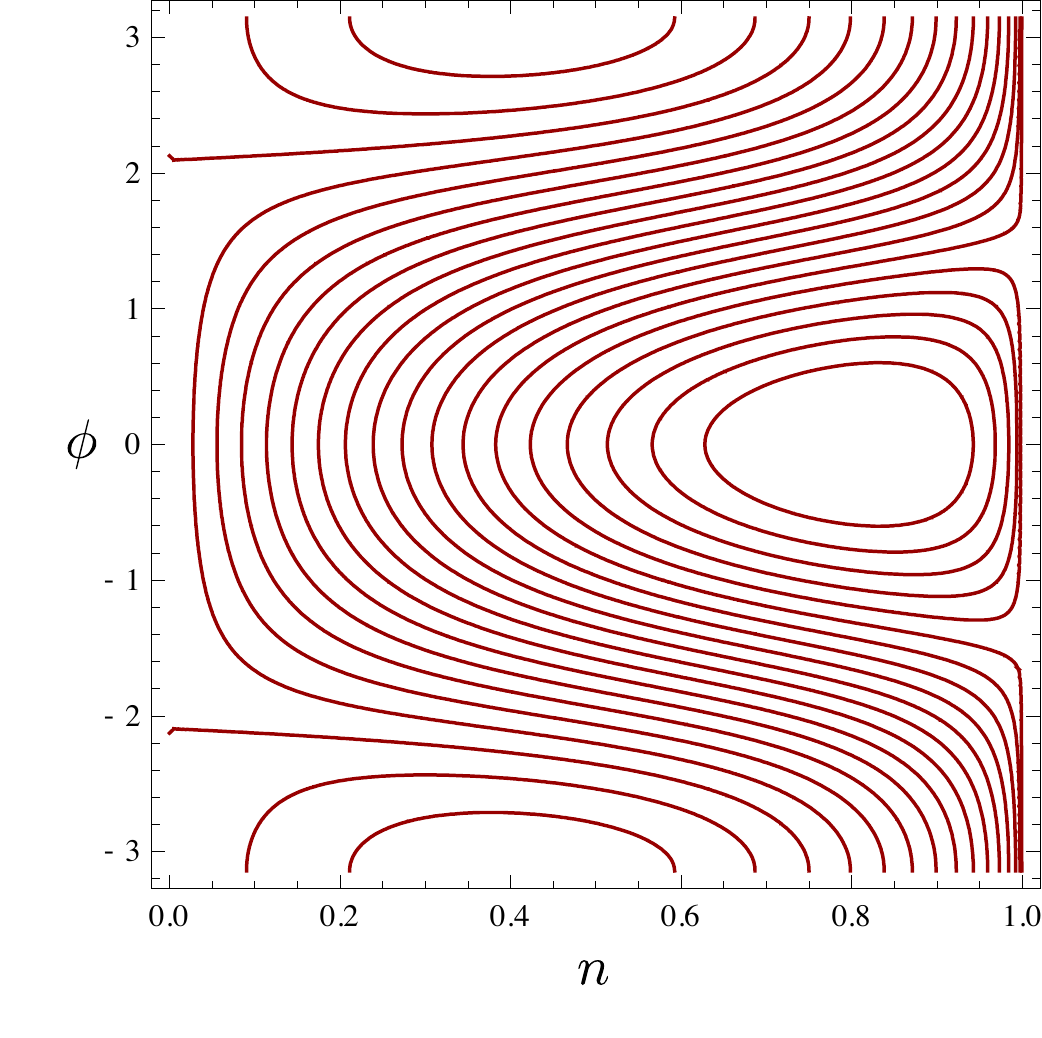}
 \caption{\label{graph3-3} Trajectories of $H$ at $\gamma=1$}
\end{figure}

\begin{figure}
\includegraphics[height=2.3in]{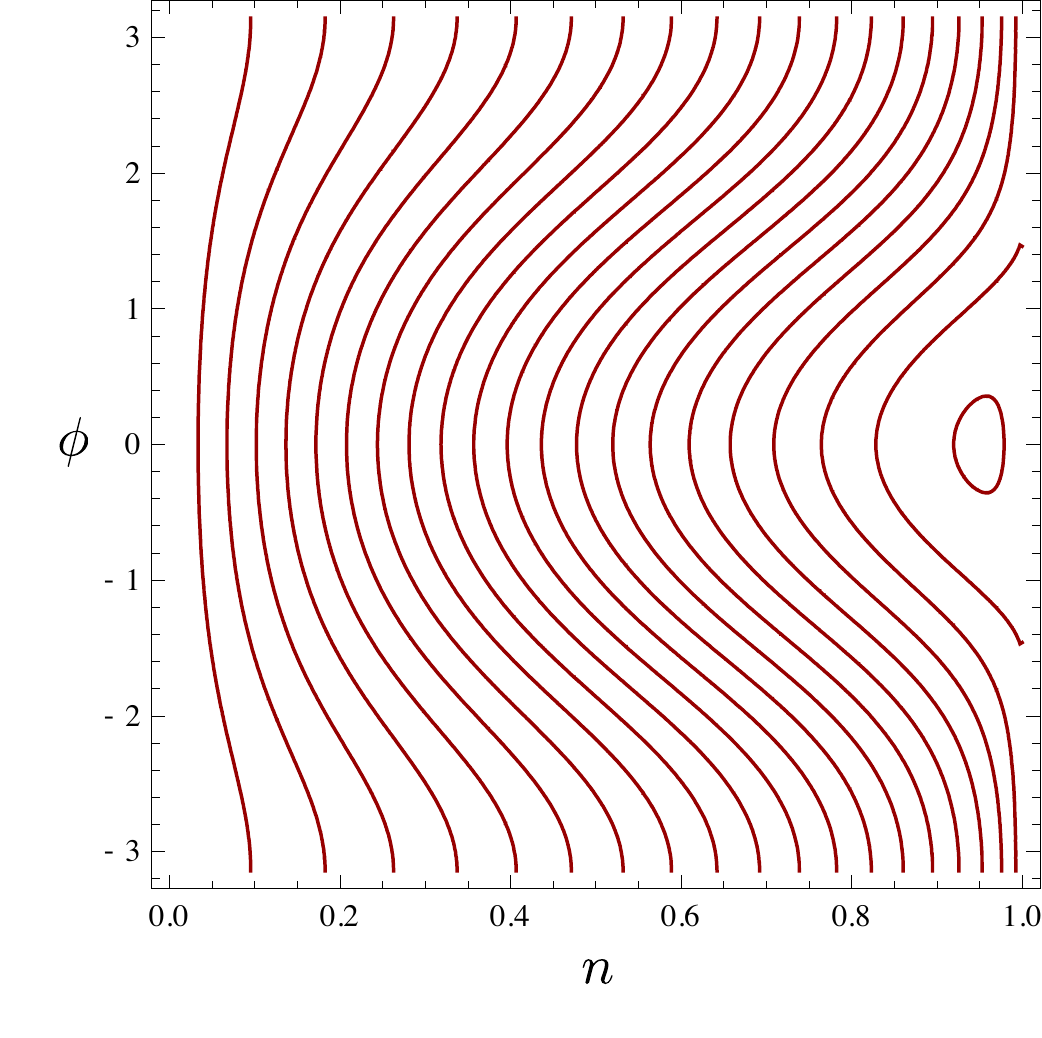}
 \caption{ \label{graph3-4} Trajectories of $H$ at $\gamma=4$.}
\end{figure}

We begin our analysis of the initial value problem by plotting lines
of constant energy $H=\mathrm{const.}$ at different fixed values of
$\gamma$. Examples are shown in Figs.~\ref{graph3-1}, \ref{graph3-2},
\ref{graph3-3} and \ref{graph3-4}.
The lines shown in these figures represent trajectories of the
Hamiltonian for $\gamma$ fixed. In our problem, $\gamma$ changes in
time. However, at small $\lambda$, $\gamma$ changes slowly. Thus the
system will follow one of the trajectories for some time before it
will slowly drift to a different trajectory due to the slowly changing
$\gamma$.

\subsection{Fixed Points}
An important role in our analysis is played by the fixed points of the
Hamiltonian system~\cite{Niu2002,Pazy2006}. This
is where the trajectory consists of a single point ($\gamma$ still
kept fixed.) They are found by solving
\begin{equation}
\frac{\d H}{\d\phi}=0, \ \frac{\d H}{\d n}=0.
\end{equation}
There are two families of solutions $(n,\phi)$ of these equations. The
first exists for $\gamma>-2$ and has $\phi=0$, $(n,\phi) \equiv
(n_1(\gamma),0)$, where
\begin{equation} \label{eq:47a}
n_1(\gamma)=\frac{12-\gamma^2+\gamma \sqrt{12+\gamma^2}}{18}
.
\end{equation}
This solution appears at $n_1(2)=0$ at $\gamma=-2$. As $\gamma$ is
increased above $-2$, $n_1(\gamma)$ starts increasing until it
asymptotically reaches $\lim_{\gamma\to \infty} n_1(\gamma)=1$.

The other solution has $\phi=\pi$, and requires $\gamma<2$. It is
given by
\begin{equation} \label{eq:48}
n_2(\gamma)=\frac{12-\gamma^2-\gamma \sqrt{12+\gamma^2}}{18}
.
\end{equation}
This solution appears at $n_2(2)=0$. As $\gamma$ is decreased below
$2$, $n_2(\gamma)$ starts increasing and asymptotes as
$\lim_{\gamma\to -\infty} n_2(\gamma)=1$.

We thus arrive at the following picture. Initially the number of
bosons is given by $n=n_0 \sim 1/N$, at undetermined $\phi$. The
coordinates $n$ and $\phi$ evolve according to the approximate
Hamiltonian $H \approx -\gamma n$, where we used that at large
negative times $\gamma$ is very large to keep only the leading order
contribution in this scale. This means that the motion is initially
given by
\begin{equation} \label{eq:49}
n= \frac 1 N, \ \phi =- \gamma  t.
\end{equation}
However, as $\gamma$ is increasing, the fixed point $n_1(\gamma)$,
given by (\ref{eq:47a}),
appears at the critical value $\gamma=-2$. As $\gamma$ keeps increasing
the point moves towards larger values of $n$. At a certain
moment of time, the trajectory $(n,\phi)(t)$, will drastically
change: it will no longer perform extended  propagatation in
$\phi$-directions but start winding around the fixed
point \rfs{eq:48}. Eventually,  at some large positive $\gamma$, the
trajectory will stop winding around the fixed point and merge into a  new
extended trajectory, which at large positive $\gamma$ will take the form
\begin{equation}
  \label{eq:50}
   n=n_{b}, \ \phi=-\gamma  t.
\end{equation}
$n_b$ is the
final number of bosons we are trying to calculate.

The fixed point $n_1(\gamma)$ has an intimate relationship with the instantaneous ground state of the quantum Hamiltonian \rfs{eq:6}. This
is discussed in more details in Appendix \ref{sec:timeind}.

\subsection{Adiabatic invariants}
In the classical mechanics of slowly parameter
dependent Hamiltonians  there exist approximately conserved
`adiabatic invariants'~\cite{LL1}. Generally, adiabatic invariants
assume the form of action integrals $\sim \int p \,dq$, where the
integral is over one full revolution of the system's motion. In the
present context, the invariant is given by
\begin{equation}
  \label{eq:22}
  I = \frac 1 {2\pi} \int d\phi \, n.
\end{equation}
To further elucidate the meaning of the adiabatic invariant in our problem, let us go back to the angular variable $\theta$
according to Eqs.~\rf{eq:42}, \rf{eq:rescale}. We find
\be \label{eq:inv1}
I = \frac 1 {4 \pi} \int d\phi \left(1 - \cos \theta \right).
\ee
This is nothing but the area of a part of the surface of a unit radius sphere bounded by the trajectory
$\theta(\phi)$, where $\theta$ and $\phi$ are thought of as spherical angles, divided by $4\pi$. This geometric definition of the adiabatic
invariant implies an ambiguity which needs to be resolved. A closed trajectory on the surface of a unit sphere separates the surface into two
domains, and one needs to pick the area of one of them.

We define the adiabatic invariant as  the area (divided by $4
\pi$) of the domain on the surface of the unit sphere bounded by the
trajectory such that the trajectory, when the domain is viewed from
above, encircles the domain in the counter-clockwise direction. We
note that this definition does not always conform to the algebraic
expression \rfs{eq:22}. However, it results in the adiabatic invariant
continuous under smooth deformations of the trajectory, while
\rfs{eq:22} can be discontinuous under those transformations.

%The adiabatic invariant is defined modulo the total surface of the sphere divided by $4\pi$, or modulo $1$. Let us illustrate this
%on the following example.

%A surface of this type has a complementary surface, so that together these surfaces constitute the entire surface of the sphere. One can equally well
%define the adiabatic invariant as the area of the complementary surface, or
%\be \label{eq:altinv}
%I = \frac{1}{4\pi} \int d\phi \left(1 + \cos  \theta \right).
%\ee
%One should use either of the two expressions, \rfs{eq:inv1} or \rfs{eq:altinv}, so that the adiabatic invariant is continuous as the trajectory is smoothly %deformed.

%Imagine a very small trajectory which winds around the north pole of the sphere in the counterclockwise direction, or $\theta=\epsilon \ll 1$,
%$\phi = \alpha t$, $\alpha>0$. Using the expression \rfs{eq:inv1} one find the invariant to be $I \approx \epsilon^2/4$. Suppose one moves the trajectory (without changing its shape) so that it now winds around some point which is no longer  the north pole. \rfs{eq:inv1} still give the same expression as before. But now suppose one moves the trajectory so that it now winds around the south pole. Then \rfs{eq:inv1}
%gives $I = \epsilon^2/4-1$. To make the adiabatic invariant continuous, one needs to add $1$ to its value. This consequence of this
%will be seen below.

For trajectories such as the initial trajectory \rfs{eq:49} or the
final trajectory \rfs{eq:50} the adiabatic invariant is very easy to calculate. Initially the trajectory encloses the north pole of the sphere in the
counter-clockwise direction, to result in
\begin{align*}
  \label{eq:21}
  I_{\mathrm{initial}}\equiv I(\gamma\to -\infty)=\frac{1}{2 \pi} \int_{-\pi}^{\pi} d\phi\, n(\gamma\to -\infty) \sim \frac 1 N.
\end{align*}
At the end of the process \rfs{eq:50} implies that the trajectory now encloses the south pole in the counter-clockwise direction.
The adiabatic invariant is now given by
\begin{equation}
  \label{eq:23}
  I_{\mathrm{final}}\equiv I(\gamma\to \infty)=1-\frac 1 {2\pi} \int_{-\pi}^{\pi} d\phi\, n(\gamma\to \infty) =1-n_b.
\end{equation}
In the  limit $\lambda \rightarrow 0$, $I$ is conserved.
\begin{equation}
  \label{eq:23a}
  I_{\mathrm{final}}=1-n_b \sim \frac {1}{N}.
\end{equation}
This implies that at the end oft the process all particles become
bosons, $n_b \approx 1$ (with $1/N$ factor coming from a quantum
uncertainty irrelevant in the large $N$ limit).

As the rate $\lambda$ is increased, the adiabatic invariant starts
changing with time. In the next section, we will review the general
theory of adiabatic action changes, and then apply it to our
particular problem at hand. Readers not interested in details of the
adiabatic dynamical evolution of the system may just note the final
result of our analysis, Eq.~\eqref{eq:79}.

% The theory of how adiabatic invariants change is
% well developed. We are going to use this theory as introduced in
% Ref.~\cite{LL1} to calculate changes in $I$. Once we know the
% asymptotic value, $I_{\mathrm{final}}$, we calculate the number of
% bosons via~(\ref{eq:23}).

\subsection{The theory of approximate conservation of the adiabatic
  invariants}
The theory of approximate conservation of adiabatic invariants is a
well developed subject, described in detail in Ref.~\cite{LL1}.  One
of its main results is that the change of an adiabatic invariant
during some time dependent process is usually exponentially suppressed
at slow rate,
\begin{equation}
  \label{eq:54}
   \Delta I \sim \exp \left( - \frac{{\rm
      const}}{\lambda} \right),
\end{equation}
where $\lambda$ is the rate of the process. In fact, one may interpret
the standard Landau-Zener transition probability in terms of this
behavior.

However, the theory also states that if a singularity develops at a
certain time during the evolution, \rfs{eq:54} breaks down and gets
replaced, typically, by a power law. We are going to see that this is
indeed the case in our problem.

For the convenience of the reader, we briefly summarize the essentials
of the theory of adiabatic invariants, as discussed in chapters 49 and
50 of Ref.~\cite{LL1}. Consider a Hamiltonian $H=H(\phi,n,\gamma)$ depending
on a pair of conjugate variables $(\phi,n)$, and a slowly varying
parameter $\gamma$. For starters, assume $\gamma$ to be
constant. Since $H$ describes a system with only one degree of
freedom, we have integrability, and it is convenient to transform to
canonical, or `action-angle' variables. The action variable is defined
as
$$
I = {1\over 2\pi} \oint d\phi\, n(E,\phi,\gamma),
$$
where the integral is over a closed trajectory of the system's
evolution. Such curves are specified by a value of the conserved
energy, $E$, and, in our case, by the value of $\gamma$. We thus have
a functional relation $I=I(E,\gamma)$. Now, consider the so-called
abbreviated action
$$
S(\phi,E,\gamma) \equiv \int_{\phi_0}^\phi d\phi' \,n(E,\phi',\gamma),
$$
where the integral extends only over a certain segment of the
trajectory. The relation $I=I(E,\gamma)$ may be inverted to express
$S$ as
$$
S(\phi,E(I,\gamma),\gamma) = S(\phi,I,\gamma)
$$
as a function of the coordinate and the action variable. The
abbreviated action is the generator of a canonical transformation
$(\phi,n)\to (I,w)$ from the old coordinates to the (conserved)
action, and a conjugate angular variable. Specifically, we have the
relation $n=\partial_\phi S(\phi,I,\gamma)\big|_{I}$ and define the angular
variable as $w \equiv \partial_I S(\phi,I,\gamma)\big|_\phi$. The
equations of motion in the new variables read
\begin{align*}
  d_t I &=-{\partial H'\over \partial w},\cr
  d_t w & = {\partial H'\over \partial I},
\end{align*}
where $H'=H'(I,w,\gamma)$ is the Hamiltonian expressed in terms of new
variables. For an autonomous system ($\gamma=\mathrm{const.}$),
$H'(I,w,\gamma) = H(I,\gamma)\equiv E(I,\gamma)$ is but the old
Hamiltonian expressed in terms of the action variable. In this case,
the action is conserved, and the angle varies as $w = t\partial_I
E$. During each period of the motion, the action changes by an amount
$\Delta S = 2\pi I$. The relation $w=\partial_I S$ then implies a
change of $w$ by $2\pi$. Thus,
$$
w(t) = t {2\pi\over T}.
$$
For a non-autonomous system, classical mechanics states that
$$
H'(I,w,\gamma)= H(I,\gamma) + {\partial S\over \partial t}=H(I,\gamma)
+ {\partial S\over \partial \gamma} d_t \gamma,
$$
where $S$ on the r.h.s. must be expressed as a function of $(I,w)$
after the  parameter differentiation.
In other words,
$$
H'(I,w,\gamma) = H(I,\gamma) + \Lambda(I,w,\gamma) d_t \gamma,
$$
where
\begin{equation}
  \label{eq:26}
  \Lambda (I,w,\gamma) =  \left.{\partial S(I,\phi,\gamma)\over \partial \gamma}\right|_{\phi=\phi(I,w,\gamma)}.
\end{equation}
This then means, that the action changes, and it does so according to
\begin{equation}
  \label{eq:25}
  d_t I = -  {\partial \Lambda(I,w,\gamma) \over \partial w} d_t \gamma.
\end{equation}
% where we introduced the abbreviation
% \begin{equation}
%   \label{eq:27}
%   R(I,w,\gamma) \equiv ,
% \end{equation}
The change of the action over a finite interval of time then follows
as
\begin{equation}
  \label{eq:59}
  \Delta I = - \int dt  {\partial \Lambda(I,w,\gamma) \over \partial
  w}d_t \gamma= -  2 \lambda   \int \frac{dw}{\omega}  {\partial \Lambda(I,w,\gamma) \over \partial
  w},
\end{equation}
where
$$
\omega \equiv {2\pi \over T}
$$
is the characteristic frequency of the trajectory. For time
independent $\gamma$, the integral is over one full interval of a
$w$-periodic integrand and vanishes; it is temporal variations in
$\gamma$ that give it its finite value.
The relations above express the evolution of the action
entirely in terms of quantities that do not contain explicit time
dependence. One may argue that $\Delta I$
evaluate to Eq. (\ref{eq:54}), unless the time
dependence of $\omega^{-1}$ and $R$ contains singularities. This will
turn out to be the case in our problem.

Figs.~\ref{graph3-1}, \ref{graph3-2}, \ref{graph3-3} and
\ref{graph3-4} show examples of trajectories that enter the
computation of action integrals for our dynamical system.  (Each of these trajectories
corresponds to a particular value of $I$ while $w$ parameterizes the
trajectory.) In computing the above integrals, we will then be met with the following scenario: consider a
trajectory, close to the left boundary of phase space and at an
initial value  $\gamma \ll 0$. Initially, that trajectory will be
'open', i.e. its angular variable will periodically run from $0$ to
$2\pi$, at moderately changing $n$. To the trajectory we may assign an
instantaneous value of energy $H(n,\phi,\gamma)$, which changes on
account of increasing $\gamma$. As $\gamma$ approaches the value
$-2$ from below, we run into a singularity: at a certain value
$\gamma=-2+\epsilon_0$, $0<\epsilon_0 \ll 1$, the energy vanishes,
$H(n,\phi,-2+\epsilon_0)=0$. That point marks a drastic
change in the topology of the trajectory, it changes from open to
closed. See Fig.~\ref{graph3-2} for an example of a closed trajectory,
and a `critical' trajectory with $H=0$. The latter begins and ends at
$n=0$, and has the form of a horizontally aligned horseshoe. The
passage time through the critical trajectory is infinite, and it is
logarithmically divergent on the closed trajectories for
$\epsilon\searrow \epsilon_0$ (in contrast, it can be checked that
another horseshoe-shaped trajectory, which starts and ends at $n=1$ and can be seen on Fig.~\ref{graph3-4}, is not critical and its passage takes finite amount of time). The action integral receives its
dominant contribution from a
range of $\epsilon$-values above $\epsilon_0$.

To
describe this situation algebraically, we expand the Hamiltonian \rfs{eq:43} for small $\epsilon$, $n$ and $\phi$, where
$$
\gamma=-2+\epsilon
$$
 as
$$
H = n \left( \phi^2-\epsilon+ n \right).
$$
The singular trajectory is given by $H=0$, or $n = \epsilon-\phi^2$.
Its frequency is zero (the period is infinity) and its adiabatic
invariant is given by
\begin{equation}
  \label{eq:iniinv} I = \frac{1}{2\pi}
\int_{-\sqrt{\epsilon}}^{\sqrt{\epsilon}} d\phi \left( \epsilon -
  \phi^2 \right) = \frac{2}{3\pi} \epsilon^{\frac 3 2}.
\end{equation}
 The
trajectories inside the singular trajectory have $H<0$ (one of these
is shown in Fig.~\ref{graph3-2}). On these closed trajectories, $n$ is
a two valued function of $\phi$, where the two branches $n_{\pm}(\phi)
= {1\over 2}\left(\epsilon-\phi^2 \pm \left(\left( \phi^2-\epsilon
    \right)^2+4H\right)^{1/2}\right)$ represent the right and left
bending arc as $\phi$ varies in $ -\sqrt{\epsilon - 2 \sqrt{-H}} \le
\phi \le \sqrt{\epsilon - 2 \sqrt{-H}}$.  The action of these curves
equals the area enclosed by the trajectories, divided by $2\pi$, or
\begin{eqnarray}
\label{eq:24}
I &=& \frac{1}{2\pi} \int_{-\sqrt{\epsilon-2\sqrt{-H}}}^{\sqrt{\epsilon-2\sqrt{-H}}}
d\phi \sqrt{\left(\phi^2-\epsilon \right)^2+4 H} \approx \cr &&
{\epsilon^{3/2}\over 2\pi} \left(\frac 4 3
- \frac{H}{\epsilon^2} \left( \log \left[ -\frac {H}{16 \epsilon^2} \right]-1 \right)\right).
\end{eqnarray}
This expression is approximate, valid for small $ \left| H \right| \ll
\epsilon^2$.  The corresponding frequencies are given by
\begin{align}
\label{eq:28}
  \omega= \frac{\partial H}{\partial I} = -\frac{2 \pi \sqrt{\epsilon}}{\log \left(-\frac{H}{16 \epsilon^2} \right)}.
\end{align}
These frequencies indeed vanish logarithmically as $H\nearrow 0$
approaches zero, confirming that the $H=0$ trajectory is critical.

The mechanical system we need to study starts its evolution at
$t\rightarrow -\infty$ by moving according to
\rfs{eq:49}. Subsequently, at a time where $\gamma$ is slightly larger
than $-2$, the system crosses the singular trajectory and starts
moving around the fixed point \rfs{eq:48}. It is at this time that the
adiabatic invariant receives most of its increase.

We estimate the increase by first calculating $S$ for the part of the
trajectory represented by $n_-(\phi)$ where $\phi<0$
\begin{align*}
  S={1\over 2}\int_0^\phi d\phi' \left(\epsilon-\phi'^2-\sqrt{\left(\phi'^2-\epsilon \right)^2+4 H}\right).
\end{align*}
In this formula, $H=H(I)$ must be understood as a function of $I$,
which can be found by inversion of (\ref{eq:24}).  We then calculate
$\Lambda$ according to (\ref{eq:26}):
\begin{align*}
  \Lambda = \frac 1 2 \int_0^\phi d\phi' \left[ 1- \frac{\epsilon-\phi'^2+2 \frac{\partial H}{\partial \epsilon}}{\sqrt{\left( \epsilon-\phi'^2\right)^2 + 4 H}}
\right].
\end{align*}
where $\phi$ has to be substituted by $\phi=\phi(I,w,\lambda)$ after
the differentiation. We now use this result to compute the time
variation of the action according to (\ref{eq:25}). Using that the
$w$-dependence of $\Lambda$ is in $\phi(w,I,\lambda)$, and that $d_t
\gamma = 2\lambda$, we have
$$
d_t I = - 2\lambda {\partial\Lambda\over \partial \phi} {\partial\phi
  \over \partial w} =
- \lambda \left[ 1- \frac{\epsilon-\phi^2+2 \frac{\partial
       H}{\partial \epsilon}}{\sqrt{\left( \epsilon-\phi^2\right)^2 +
       4 H}}\right] {\partial\phi\over \partial w}.
$$
Eventually, this quantity has to be integrated over time, and to do
so, we need information on the time dependence of $\phi(I,w,\gamma) =
\phi(I,\omega t,\gamma)$. The equations of motion tell us
\begin{equation}
  \label{eq:eqonfi}
d_t \phi = \partial_n H= - \sqrt{ \left(\epsilon - \phi^2 \right)^2 + 4
  H}.
\end{equation}
We next assume that during the time when the invariant $I$ receives
most of its increase, $\phi$ takes values such that the r.h.s. of this
equation vanishes. The validity of this presumption will be checked in
the end of the section. This implies
\begin{equation}
  \label{eq:condonfi}  \phi \sim - \sqrt{\epsilon}.
\end{equation}
Indeed, everything done so far implies $ \left| H \right| \ll
\epsilon^2$ and so the right hand of \rfs{eq:eqonfi} approximately
vanishes if \rfs{eq:condonfi} holds.

Using this stationarity condition, and
 $\partial_w \phi = \omega^{-1}d_t \phi$ we obtain
$$
d_t I \simeq {2 \lambda \over \omega}
  \left( \sqrt{-H} - \partial_\epsilon H \right).
$$
In this equation $H=H(I,\epsilon)\ll \epsilon^2$ is
implicitly defined by (\ref{eq:24}). This gives
$$
{\partial H \over \partial \epsilon} = - {\partial I\over \partial
  \epsilon} \left({\partial I \over \partial H}\right)^{-1}\simeq
{2\epsilon \over\ln\left(- {H \over 16 \epsilon^2}\right)}
,
$$
a term that is larger than the $\sqrt{-H}$ contribution to $d_t
I$. Using (\ref{eq:28}), we arrive at the estimate
$$
d_t I \simeq {2\lambda \epsilon^{1/2}\over \pi}.
$$
Using that $\epsilon = 2 + 2\lambda t$, we integrate this expression
over $\epsilon$ from $\epsilon=\epsilon_0$, remembering that $I=2
\epsilon_0^{3/2}/(3 \pi)$ when $\epsilon=\epsilon_0$, to find
\begin{equation}
  \label{eq:invinc} I \simeq {2\over 3\pi} \epsilon^{3/2}.
\end{equation}
This expression looks formally identical to \rfs{eq:iniinv}, but has a
different meaning.  While \rfs{eq:iniinv} is the adiabatic invariant
of the critical trajectory, \rfs{eq:invinc} signifies that this
expression remains approximately true even at later times when the
system no longer follows the critical trajectory.

% \be R = - \frac 1 2 \left[ 1+ \frac{\epsilon-\phi^2+2 \frac{\partial
%       H}{\partial \epsilon}}{\sqrt{\left( \epsilon- \phi^2 \right)^2
%       + 4 H}} \right] \frac{\partial \phi}{\partial w}.  \ee Using
% \be \frac{\partial \phi}{\partial t} = \sqrt{ \left(\epsilon -
%     \phi^2 \right)^2 + 4 H}, \ee which follows from the equations of
% motion, we find \be R \approx \left( 2 \sqrt{-H} + \frac{\partial
%     H}{\partial \epsilon} \right) \frac{ \log \left[-\frac{16
%       \epsilon^2}{H} \right]}{2 \pi \sqrt{\epsilon}}, \ee which is
% only true at $H$ close to zero. This finally gives \be R \approx
% \frac{\sqrt{\epsilon}}{\pi}.  \ee It follows that \be \dot I = 2
% \lambda \frac{\sqrt{\epsilon}}{\pi}, \ee or \be \label{eq:76} I =
% \frac{2 \epsilon^{\frac 3 2}}{3 \pi}.  \ee In other words, the
% adiabatic invariant approximately increases as if the system stays
% in the critical trajectory.

We finally need to find the maximum value $\epsilon\equiv \epsilon^\ast$
for which \rfs{eq:invinc} still holds (at even higher values of
$\epsilon$, the adiabatic invariant will stop growing and just
oscillate about its average value). The criterion we will use to
determine that value reads $w=\pi$, i.e. we demand that the system
proceeded along half a period of the critical trajectory. At larger
instances of time ($\epsilon$) the transit into the domain of
oscillatory motion and no further systematic action increase has taken
place. In practice, the fixation of $\epsilon^\ast$ turns out to be
somewhat tedious, and we have relegated it into appendix
\ref{sec:epsilon}. As a result we obtain
\begin{equation}
  \label{eq:29}
  \epsilon^* \sim \left(\frac 3 2 \lambda \log N
\right)^{\frac 2 3}.
\end{equation}

Finally, with the help of \rfs{eq:23a} this leads to our final answer
\begin{equation}
  \label{eq:79}
  n_b \simeq   N \left( 1 - \frac{\lambda}{\pi g^2} \log N \right),
\end{equation}
where we have re-introduced the coupling strength $g$ and returned to
the original (unrescaled) $n_b$. Note that in the derivation above we
relied only on the fact that the unrescaled $n_0\sim 1$ due to quantum
fluctuations. If we start from the state where $n_0$ is non-zero (but
smaller than $N$) due to initial occupancy of the molecular state the
result (\ref{eq:79}) remains valid with the only difference that the
argument of the logarithm should be changed to $N/n_0$. This is
qualitatively different from the results of Ref.~\cite{Pazy2006} where
an instant crossover to a $\sim \lambda^{1/3}$ power law the moment
$n_0 \gtrsim 1/N$ was predicted.

%Up to the
%prefactor {\tt substantiate} this agrees with Eq. (26) of
%Ref. ~\cite{Pazy2006}. Importantly, however, we predict a very different
%application range of the result: the calculation above entails that
%Eq. (\ref{eq:79}) holds for the entire range of driving parameters
%(\ref{eq:limit}). Below, we will present compelling numerical evidence
%in support of this estimate. {\tt do we, actually?} In contrast,
%Ref.~\cite{Pazy2006} limits the validity of (\ref{eq:79}) to vanishingly small
%initial occupancies of the boson state. For generic initial values, a
%power law $1-n_b\sim \lambda^{1/3}$ is predicted. Our direct numerics,
%the TWA scheme, and the analytical calculation above indicate that
%this prediction is incorrect.

\section{TWA and comparison with numerical simulations}
\label{sec:numerics}

In this section, we will apply a combination of numerical
diagonalization and the TWA to obtain accurate results for both, the
mean value of $n_b$ and its distribution.  For moderate values of $N$,
the Schr\"odinger equations for \rfs{eq:6} can be solved
directly. Indeed, the Hilbert space corresponding to that Hamiltonian,
in the sector $S_z + b^\dagger b = N/2$, has only $N+1$ states. Thus
the Hamiltonian reduces to a $(N+1)$-dimensional matrix. This matrix
can be diagonalized at reasonable numerical cost up to values $N
\lesssim 10^3$ (see Appendix~\ref{sec:timeind} for an example of how
this procedure can be set up).

At larger values of $N$, we simulate the classical equations of
motion, with initial drawn from the quantum Wigner distribution
Eq.~(\ref{wig_func}) -- the TWA. In this way, we may obtain results for
significantly larger values of $N$.
In the following  we will apply both methods interchangeably.
However, before doing so, let us first test the accuracy of TWA by
comparison to numerical diagonalization at values of $N$ where both
methods are applicable.

\subsection{TWA vs direct diagonalization}
\label{sec:twa-vs-direct}

\begin{figure}[h]
  \centering
  \includegraphics[width=3.3in]{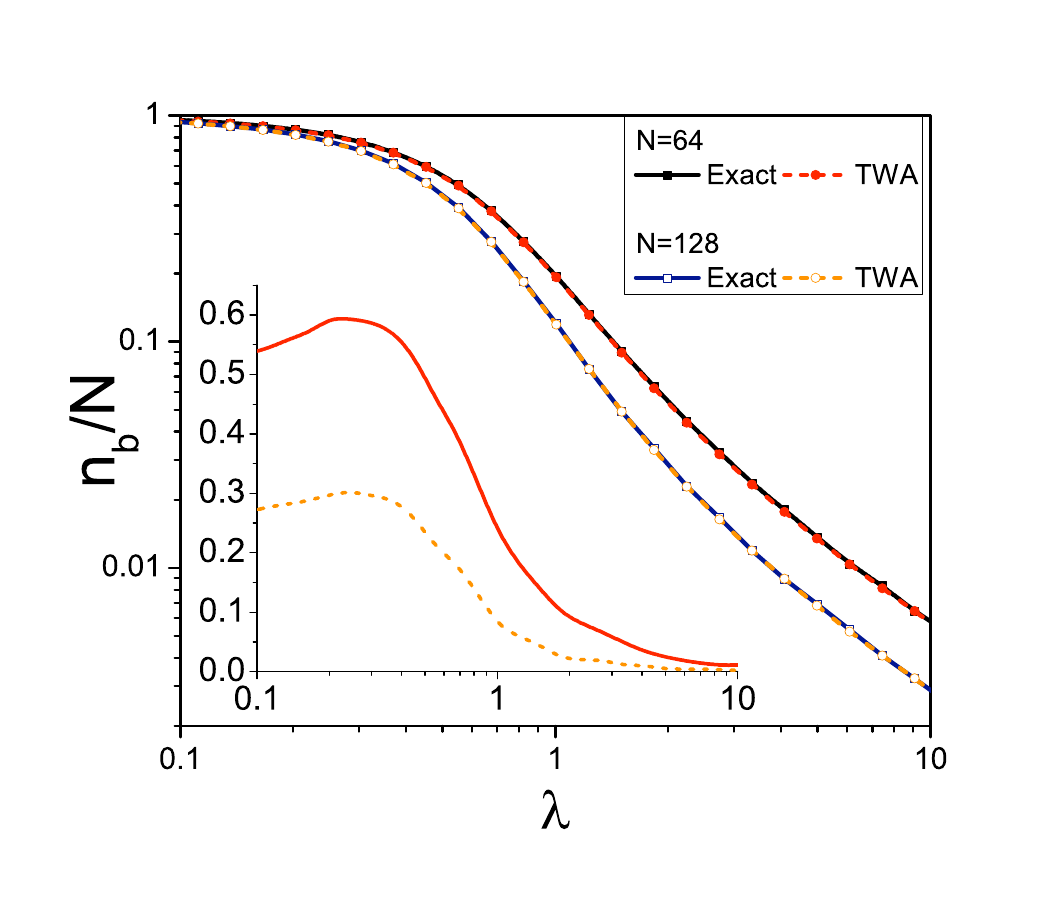}
  \caption{\label{twa_check} Dependence of the scaled number of bosons
    $n_b/N$ as a function of the rate $\lambda$. The two pairs of
    curves correspond to $N\equiv 2S=64$ and $N=128$. The solid lines
    represent the exact solution while the dashed lines are the
    semiclassical TWA. The inset shows the difference between exact
    and semiclassical results ($N=64$ red solid line and $N=128$
    orange dashed line) multiplied by a factor 100. }
\end{figure}

In section \ref{sec:twa} we
have argued that the exactness of the TWA in the linear regime entails
its applicability to the whole range of driving parameters. To back up
this claim, let us compare TWA results to those obtained by numerical
diagonalization.  In Fig.~\ref{twa_check} we show $\langle n_b\rangle
(\lambda)$ for two values of $N=2S$: $N=64$ and $N=128$. The solid and
dashed lines represent the exact and TWA solutions,
respectively. There is no visible difference between them. To
demonstrate that TWA is not actually exact we show in the inset the
difference between TWA and numerical results multiplied by a factor of
$100$. Clearly for $N=128$ the accuracy of TWA is better which signals
that as $N$ increases TWA gives results of increasing precision.

% Here we will
% verify the accuracy of TWA comparing exact and TWA numerics for
% moderately large $N$. Notice that in principle the TWA is guaranteed
% to work at short times. However, in practice it can be very accurate
% even for very long processes (see Ref.~\cite{ap_twa}) for the
% details. For this problem one can expect that TWA is accurate in the
% whole range of times. Indeed at short times where the linearized
% approximation is valid TWA is exact. By the time the nonlinearities
% become important $n_b$ becomes sufficiently large and can be treated
% classically. These considerations suggest that the regime of
% applicability of TWA should span wide range of rates $\lambda$.

\begin{figure}[h]
  \centering
  \includegraphics[width=3.3in]{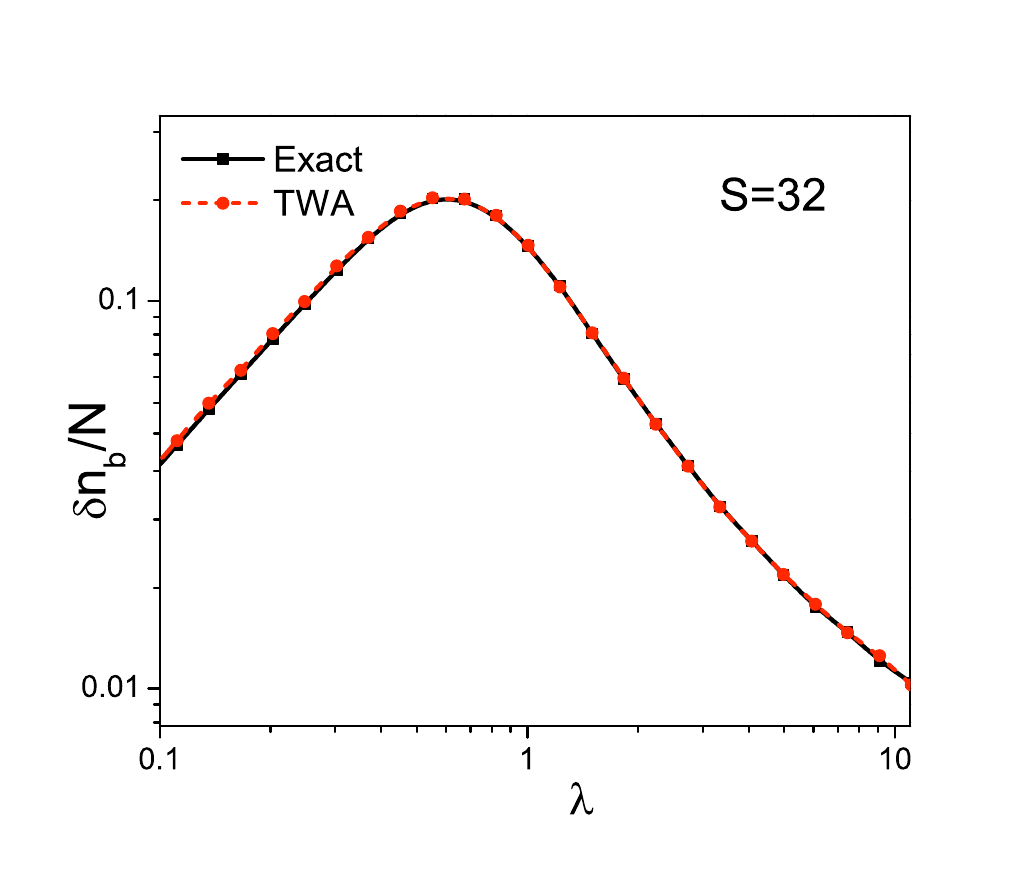}
  \caption{\label{twa_check_var} Dependence of the scaled variance of
    the number of bosons $\delta n_b/N$ after the process as a
    function of the rate $\lambda$ for $N=64$. The two curves
    represent the exact and the TWA results.}
\end{figure}
In Fig.~\ref{twa_check_var} we plot the relative number fluctuations
$\delta n_b/N=\sqrt{\langle n_b^2\rangle-\langle n_b\rangle^2}/N$ as a
function of $\lambda$ for $N=64$. As with the dependence $n_b(\lambda)$
the difference between TWA and exact results is minuscule and it
vanishes fast as $N$ gets larger. These results make us confident
that TWA represents is high precision method which becomes
asymptotically exact in the limit $N\to \infty$.

\subsection{Combined application of TWA and direct diagonalization}
\label{sec:comb-appl-twa}

In the following we will analyse our system by combined application of
the two numerical methods. We will not indicate which method was used for which
particular curve, unless necessary. In all numerical simulations we set $g=1$. (This can
be always achieved by the rescaling $\lambda\to \lambda g^2$.) Unless
stated differently, the notation $n_b(\lambda)$ refers to the mean
value of the number of bosons.

\begin{figure}[h]
  \centering
  \includegraphics[width=3.3in]{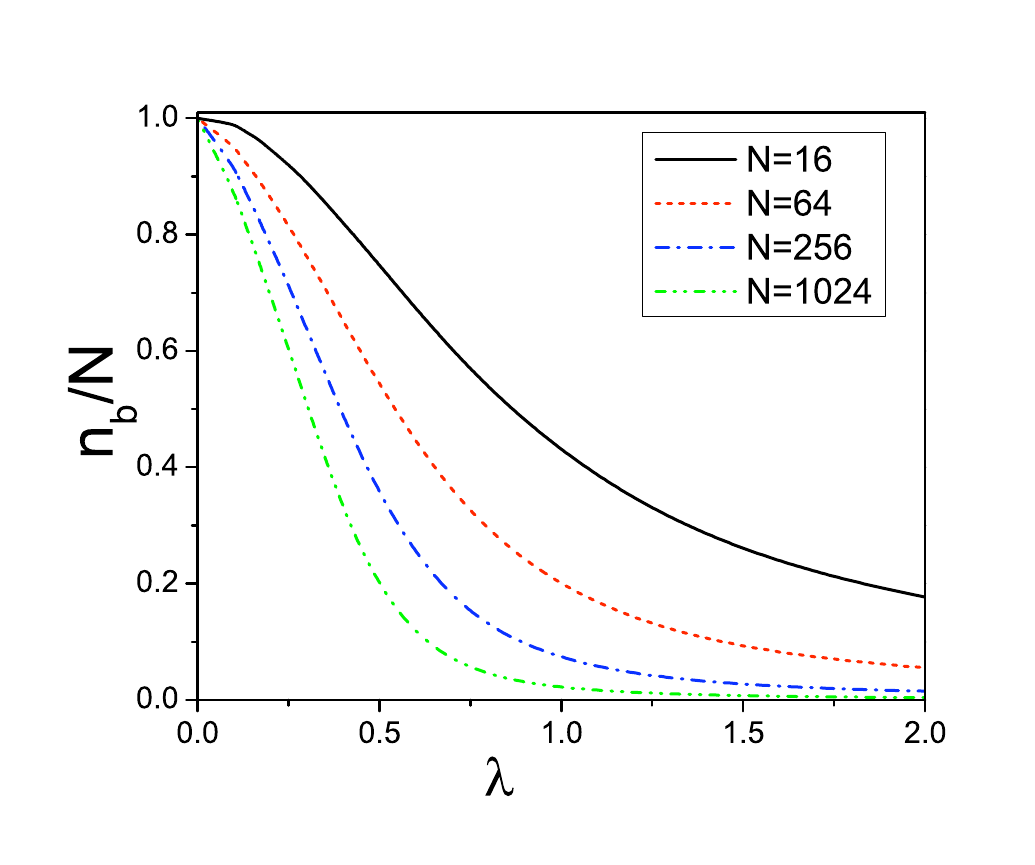}
  \caption{\label{fig:num1} Number of remaining bosons $n_b$ vs the
    rate $\lambda$ for different values of $N$. It is clear that as
    $N$ increases the dependence goes from exponential form
    characteristic for Landau-Zener process to the linear one
    predicted analytically (see Eq.~(\ref{eq:79})) }
\end{figure}

In Fig.~\ref{fig:num1} we plot $n_b(\lambda)$ for different values of
$N$. For small $N$ the function $n_b(\lambda)$ is qualitatively
similar to an exponential form: $n_b(\lambda)/N\propto 1-1/x=1-\exp(-\pi /
\lambda)$, as with the conventional two-level Landau-Zener
problem. However, as $N$ increases this exponential behavior
disappears and the dependence becomes much closer to the linear one
predicted in Eq.~(\ref{eq:79}).
\begin{figure}[h]
  \centering
  \includegraphics[width=3.3in]{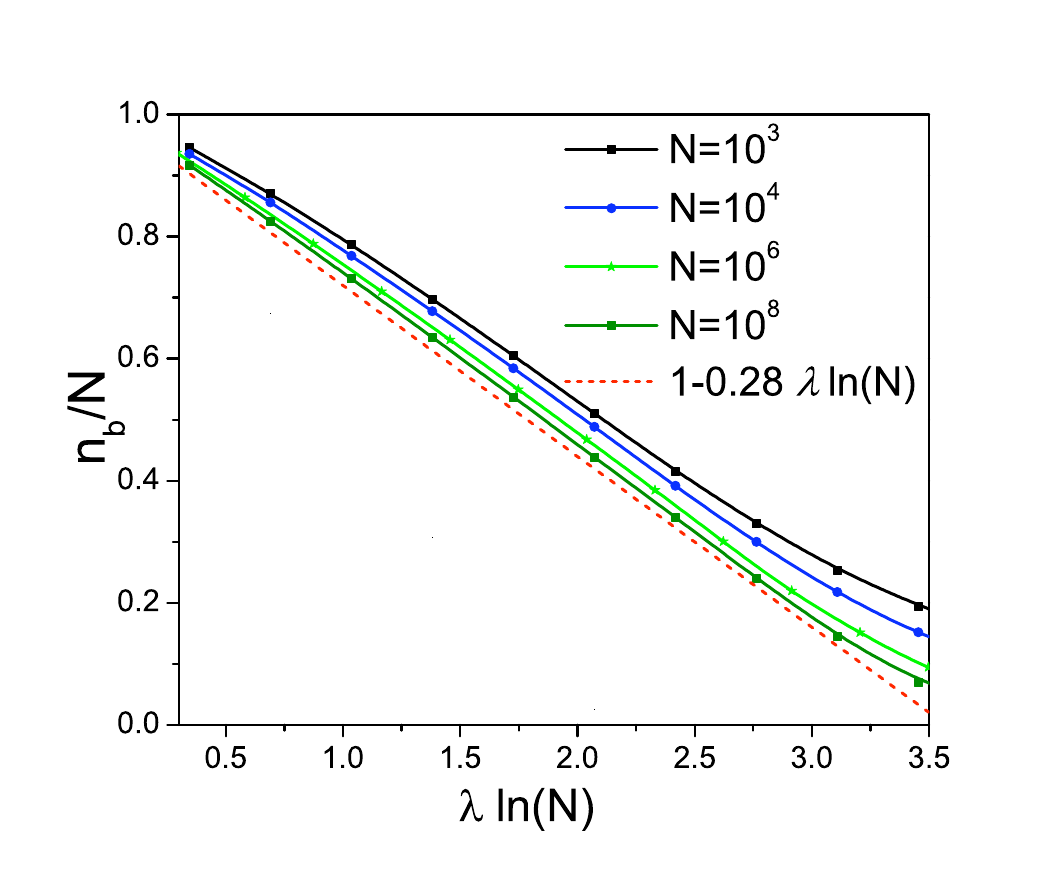}
  \caption{\label{fig:num2} Same as in Fig.~(\ref{fig:num1}) but for
    large values of $N$. Note that the horizontal axis is $\lambda\ln
    N$.  }
\end{figure}
To demonstrate the approach to linear scaling, we plot $n_b(\lambda)$
for large values of $N$ up to $N\sim 10^8$ in Fig.~\ref{fig:num2}. The
dashed line is the fit to the linear dependence: $n_b/N\approx
1-0.28\lambda\ln N$. Note that the prefactor $0.28$ is slightly
different from the $1/\pi$ in Eq.~(\ref{eq:79}). But given that the
analytic result was obtained using a series of approximations we find
this level of agreement to be satisfying.

\begin{figure}[h]
  \centering
  \includegraphics[width=3in]{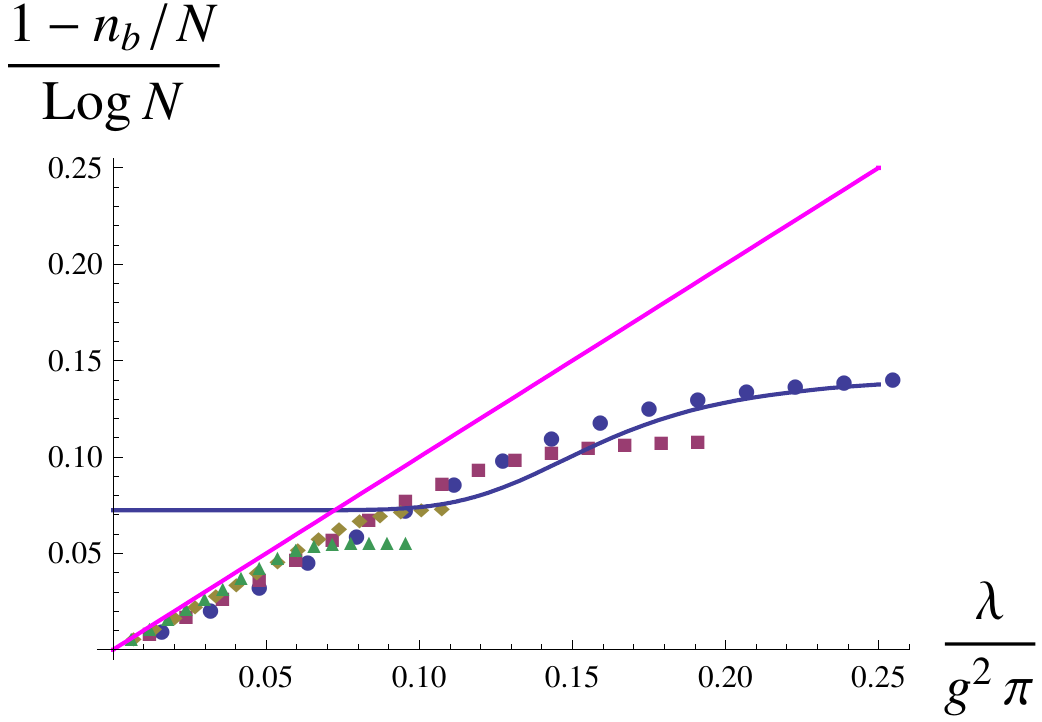}
  \caption{\label{fig:num} Number of remaining bosons $n_b$ vs the
    rate $\lambda$ for different values of $N$. Here circles represent
    $N=10^3$, squares $N=10^4$, rhombi $N=10^6$, and triangles
    $N=10^8$. The continuous lines represent our analytic solutions,
    as discussed in the text.}
\end{figure}

Fig.~\ref{fig:num} shows the results of numerical simulations at
$N=10^3$, $N=10^4$, $N=10^6$ and $N=10^8$, plotted as $(1-n_b/N)/\log N$
vs $\lambda/(\pi g^2)$. The straight line in the plot represents the
deep adiabatic limit \rfs{eq:79}, while the second continuous line
is the solution of the kinetic equation \rfs{eq:kinan} at
$N=10^3$ only.

We see that the deep adiabatic relation \rfs{eq:79} is confirmed in
the limit of slow rates. As to the solution of the kinetic equation,
while it describes well the relatively fast rates, it deviates from
the numerics at slower rates and eventually breaks down when it
saturates at nonzero value when $\lambda \rightarrow 0$.

Next, let us discuss the distribution function $P(n_b)$. In the fast
limit we have a well defined analytic prediction given by
Eq.~(\ref{eq:19})
\begin{figure}[h]
  \centering
  \includegraphics[width=3.3in]{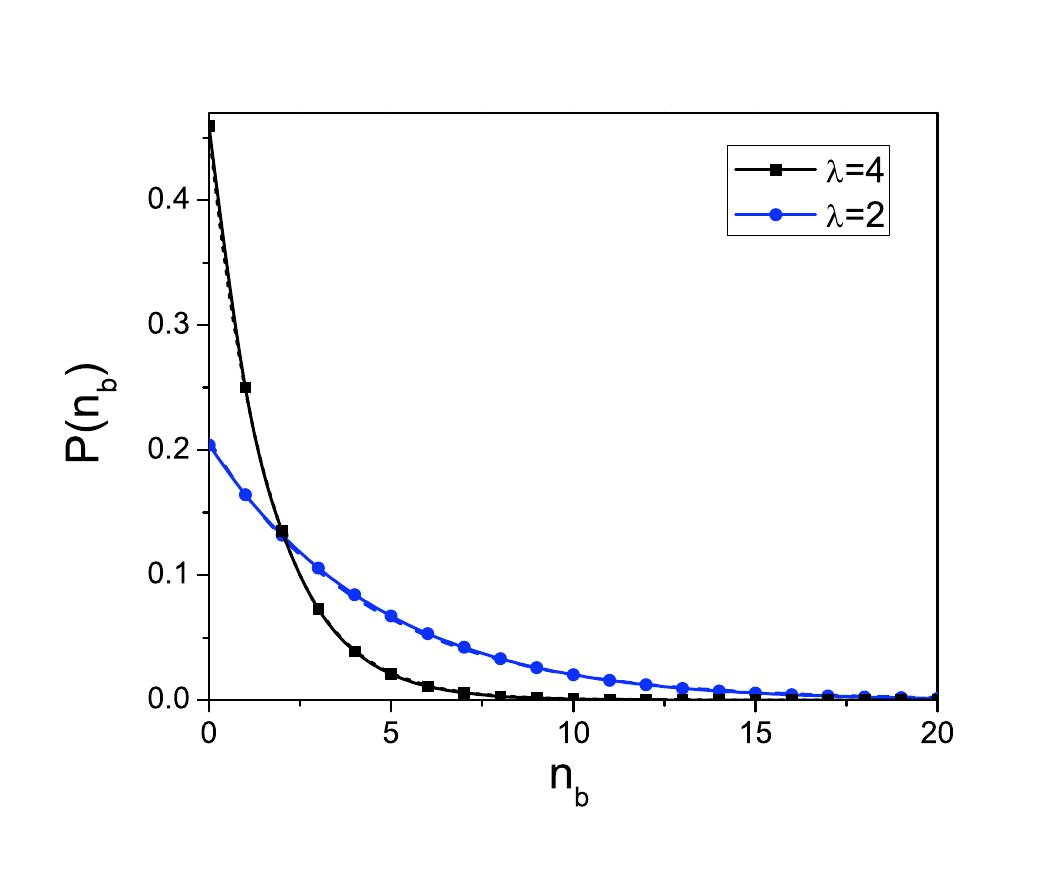}
  \caption{\label{fig:dist1} Distribution function of $n_b$ for
    $N=128$ and two values of the rate: $\lambda=2$ and
    $\lambda=4$. The distributions are discrete and the lines are the
    guide to the eye. The dashed lines, which are barely
    distinguishable from the numerically computed solid lines, are the
    analytic prediction~(\ref{eq:19}.)  }
\end{figure}
While for the slow limit we do not have an analytic prediction for the
distribution function, one can make a good ansatz based on the result
Eq.~(\ref{eq:79}). Note that this result was obtained for a particular
choice of $n_0=1$. More generally for each fixed initial value of
$n_0$ it should read: 
\be 
n_b(n_0)/N\approx 1-{\lambda\over C_1}\ln(C_2  N/n_0),
\label{eq:slow1}
\ee where $C_1$ is a constant close to $\pi$ and $C_2$ is another
constant, which in general can depend on $\lambda$. Finding this
constant is beyond accuracy of our analytic derivation. Because we
know the distribution of $n_0$: $P(n_0)=2\exp[-2n_0]$ we can
invert Eq.~(\ref{eq:slow1}) and derive the distribution function of
$n_b$:
\be
P(n_b)\approx 2{C_1\over\lambda}C_2 \mathrm
e^{-\rho}\exp\left[-2C_2 N\mathrm e^{-\rho}\right],
\label{gumbel}
\ee
\begin{displaymath}
\rho={C_1\over\lambda}\left(1-{n_b\over N}\right).
\end{displaymath}
The function $P(n_b)$ above describes the Gumbel distribution, which
often appears in the context of the extreme value statistics.

\begin{figure}[h]
  \centering
  \includegraphics[width=3.3in]{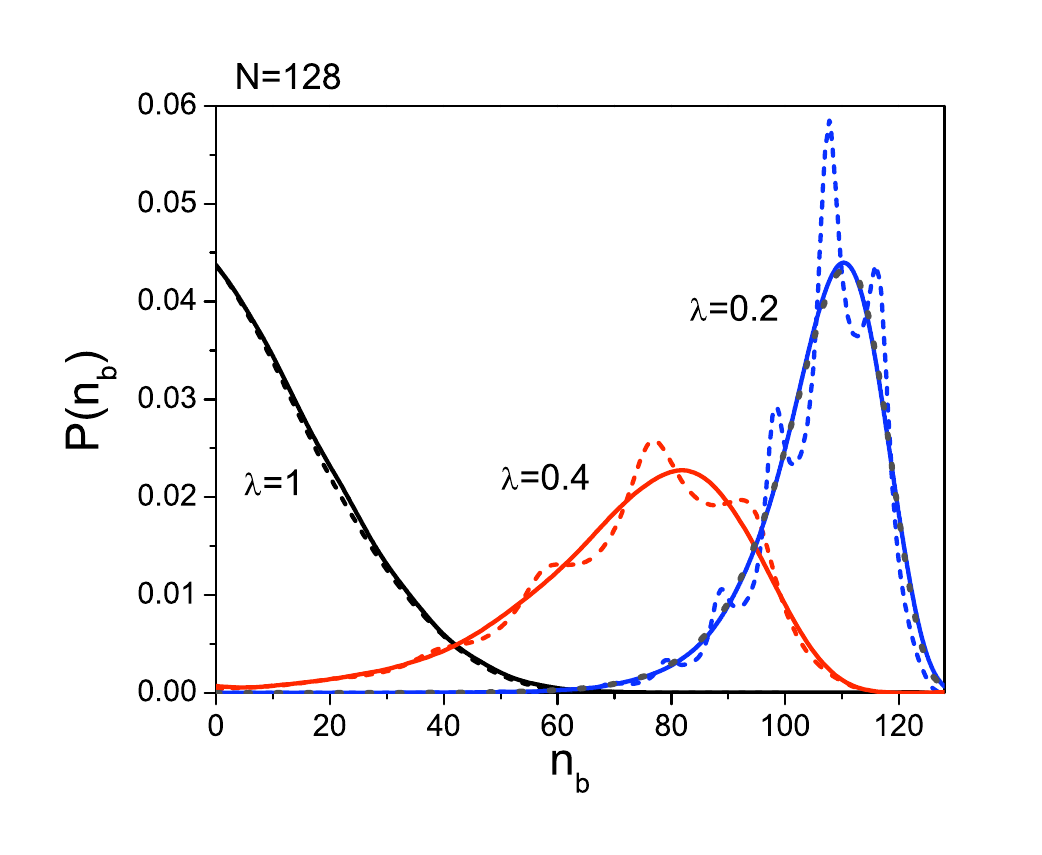}
  \caption{\label{fig:dist2} Distribution function of $n_b$ for
    $N=128$ at $\lambda=1,\; 0.4,\;0,2$. The solid lines are the
    result of the semiclassical TWA approximation and the dashed lines
    are the distributions obtained by exact diagonalization. The
    dotted grey line (which almost coincides with the solid blue line)
    is the fit to the Gumbel distribution (\ref{gumbel}) with
    $C_1\approx 3.05$ and $C_2\approx 0.03$.)  }
\end{figure}

We plot the distribution function $P(n_b)$ in
Fig.~\ref{fig:dist2}. The solid lines are the result of the
TWA approximation and the dashed lines are the exact
result. At slow rates the distribution
function becomes highly asymmetric. At $\lambda=0.2$ it is very well
fitted by the Gumbel distribution~(\ref{gumbel}) (grey dotted line)
with $C_1\approx 3.05$ (instead of $\pi$) and $C_2\approx 0.03$. The
fact that $C_2$ is so small probably indicates that this constant
depends on $\lambda$. We checked that the fit also works well for
$N=256$ with the same constants $C_1$ and $C_2$. From
Fig~\ref{fig:dist2} it is apparent that the semiclassical
approximation correctly predicts the Gumbel-like shape of the
distribution function at small $\lambda$ but misses the oscillations
on top of this shape. These
oscillations have a minor effect on the expectation value of $n_b$
and on its fluctuations. However, they are not negligible
and they do not vanish if we increase $N$ (at least they persist up to
$N=512$). We are not sure what the origin of these oscillations is. It
is interesting to point out that a similar interpolation of the
distribution function from the exponential to the Gumbel form was
found in the completely different context of interference between two
independent Luttinger liquids~\cite{gadp}.

\subsection{Discussion}
\label{sec:disc-comp-prev}

The result \eqref{eq:79} implies that the thermodynamic limit of our
system never behaves adiabatically: no matter how small the driving,
in the limit $N\to \infty$, the adiabatic limit becomes elusive. Such
type of behavior has been found earlier~\cite{np} for more general
low-dimensional bosonic systems near instabilities such as second
order phase transitions.  The giant quantum
fluctuations accompanying this non-adiabatic behavior result from the
inversion of the classical ground states of the participating
systems. Zero point fluctuations
are required to trigger a macroscopic inversion of state occupancies
and the  stochastic nature of this initiation process  
generates vast fluctuations at later stages. 

Finally, it is worthwhile putting the results above in a more general
context: a number of previous works~\cite{ap_05, zurek_05,
    dziarmaga_05, sengupta_08, sengupta_08a,dutta, fazio} have addressed the slow
  dynamics of low-dimensional {\it critical} systems. In all these cases it
  has been argued that the number of excitations produced upon  slow
  driving
  scales as a power law of the rate (as opposed to the the exponential
  dependence Eq.~(\ref{lz}) for the Landau-Zener problem.) The
  exponent of this power law is related to the critical exponents
  characterizing the phase transition~\cite{ap_05,
    zurek_05} under consideration.
  % In Ref.~\cite{np} it has in fact been argued that the power law
  % dependence of various thermodynamic observables is generic for all
  % gapless systems.
In low dimensions with their relatively higher density of low
  energy states there is a stronger tendency for  smaller powers,
  i.e. less adiabatic regimes. In a sense, our 'infinite-range
  interaction' or 'mean field' system exemplifies these structures in
  the extreme setting of a zero-dimensional system.

%   Additional tendencies to be less adiabatic arise in systems with
%   bosonic excitations there can be additional tendency to be less
%   adiabatic related to bosonic bunching~\cite{np}.

% Let us emphasize that the linear rather than exponential dependence of
% $n_b(\lambda)$ is in accord with general expectations (see
% Ref.~\cite{np}). It is the exponential dependence in the usual
% Landau-Zener problem, which is rather non-generic. The fact that there
% is no adiabatic limit at $N\to\infty$ is also consistent with one of
% the regimes suggested in Ref.~\cite{np}.  Such non-adiabatic regime
% can be realized in low-dimensional systems with bosonic excitations
% near singularities like second order phase transitions. In our case
% such singularity corresponds to the qualitative change in the shape of
% trajectories when $\lambda t$ crosses $\pm 2$ (see
% Sec.~\ref{sec:deepadiabaticlimit}). Near these singularities the
% adiabatic invariants rapidly change leading to the strong nonadiabatic
% response of the system.

\section{Application to other driving processes}
\label{sec:alt}

We spent most of this paper discussing driving processes where the
bosonic sector of the system was initially empty
% the problem represented by the
%Hamiltonians Eqs.~\rf{eq:6}, \rf{eq:5}, \rf{eq:4}, \rf{eq:17} and
%\rf{eq:hamdifbos}, where initially, in the language of Eqs.~\rf{eq:6},
%\rf{eq:5} and \rf{eq:4} there were no bosons.
(or there were no a atoms in the language of the atom-molecular
conversion systems
Eqs.~\rf{eq:17} and \rf{eq:hamdifbos}.)

For these initial conditions, quantum fluctuations are needed to jump
start boson (or molecule) production, and this is what ultimately
leads to the power law \eqref{eq:79}.
% What makes these initial conditions distinct is that they require an
% initial quantum fluctuation in the number of bosons to jump-start
% their production. This is what subsequently leads to the condition
% \rfs{eq:3} for the substantial conversion of the particles into
% bosons (or in the language of Eqs.~\rf{eq:17} and \rf{eq:hamdifbos}
% for the substantial atom production).
We here briefly discuss the `inverse' problem, where initially all $N$
particles were bosons (or atoms). This variant was previously
discussed in Ref.~\cite{Liu2008}. What makes it different, is that no
quantum fluctuation are required to start the conversion process.
Solution of the classical equations of motion \rfs{eq:43} at inverted
driving rate, $\gamma=-2 \lambda t$, and with initial condition $n(t
\rightarrow -\infty)=N$ captures the dynamics of the process and
quantum fluctuation introduce only minor corrections.

As a result of a straightforward adaption of the discussion of section
 \ref{sec:keldysh}, we find that the number of bosons $n_b$
for \rfs{eq:6} in the infinite future behaves as
\begin{equation}
   \label{eq:inv} n_b \simeq
\frac{N}{2-e^{-\frac{\pi g^2}{\lambda}}}.
\end{equation}
At fast
rates, this is equivalent to $n_b \simeq N e^{-\frac{\pi
    g^2}{\lambda}}, \ \lambda \gg \pi g^2$, which is but
the standard Landau-Zener prediction. As the rate slows down, below
$x= e^{\pi g^2/\lambda} \sim 1$ the Landau-Zener
prediction breaks down. (Notice that the initial [linear] regime in
the direct problem was much more robust.)

Taken at a face value, \rfs{eq:inv} predicts that at slow driving rates
$n_b$ saturates at $1/2$. This cannot be correct since \rfs{eq:inv}
must break down at slower rates. A detailed analysis involving the
quasiclassical Hamiltonian \rfs{eq:43} with the initial condition
$n_0=1$, gives 
\be 
n_b/N \simeq 0.057 \, \lambda, 
\ee 
where the constant of proportionality $0.057$ was determined numerically from the linear fit (note that unlike that in \rfs{eq:79}, it is $N$-independent).
\begin{figure}[h]
  \centering
  \includegraphics[width=3.3in]{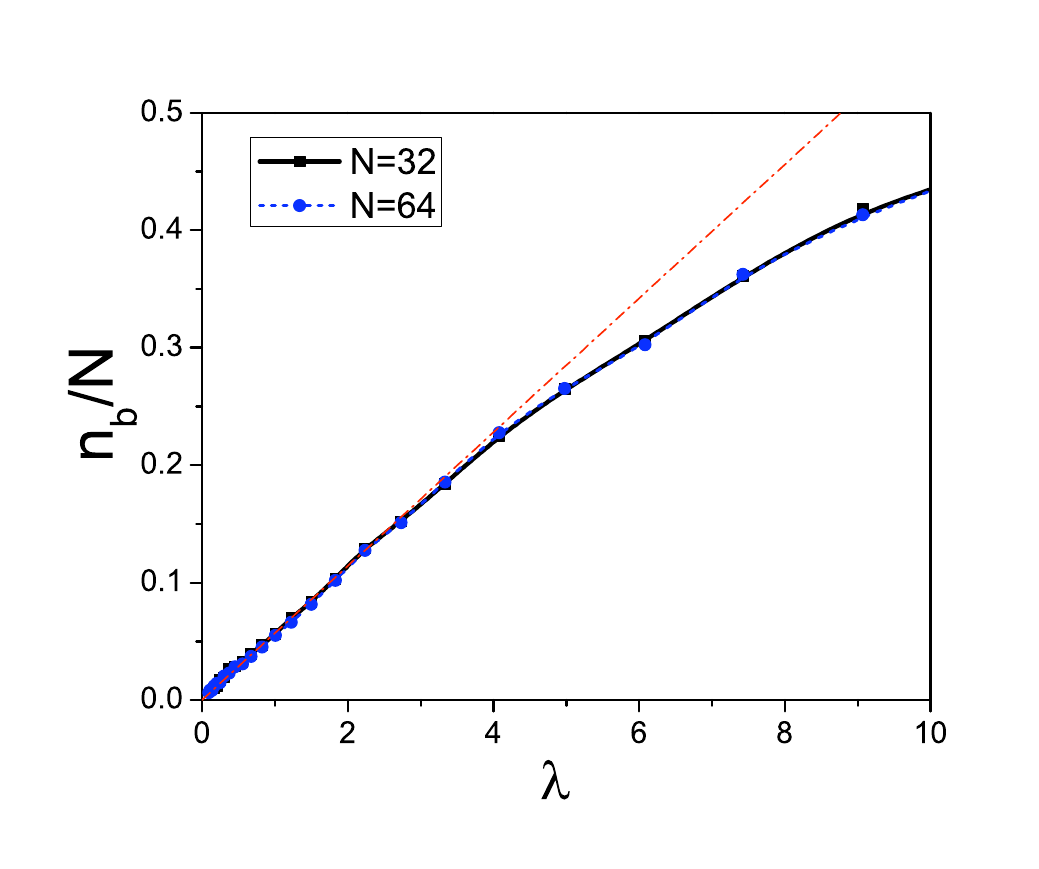}
  \caption{\label{fig:bos} The number of remaining molecules vs. $\lambda$ for the inverse problem, where we start from the full molecular condensate and no atoms. Note that the results for $N=32$ and $N=64$ almost identically coincide. The thin red line represents the linear fit at slow rates: $n_b/N\approx 0.057\lambda$.)
    }
\end{figure}
In Fig.~\ref{fig:bos} we show numerical data for the dependence
$n_b(\lambda)$ for the inverse problem for two values of $N=32$ and
$N=64$.  The thin red line represents the linear fit at slow rates:
$n_b/N\approx 0.057\lambda$, and the numerical data approaches this
asymptotic in an $N$-independent manner.

Thus the number of remaining bosons, a measure of deviation of
adiabaticity, remains proportional to the rate at slow rates, as
opposed to the exponential suppression of the standard Landau-Zener
process.  Even though one approaches the adiabatic regime when the
Landau-Zener parameter is close to $1$, the deviation from
adiabaticity is still significant and much larger than in the standard
problem. This may have significant implications for the experiments
such as those reported in Refs.~\cite{Hodby:2005uf,Papp:2006fx},
provided the initial state is chosen to be atoms and one is interested
in the molecule production. We also note that, as far as we can see,
this point was not discussed in Ref.~\cite{Liu2008}.

In some applications it may be important ~\cite{Sun:2008qm} to explore
the sector of the theory where the total number of particles is much
smaller than $N$ (i.e. much smaller than the number of two-level
systems in \rfs{eq:5}, or smaller than the number of fermionic pair
levels in \rfs{eq:4}.) Once more, it is easiest to analyze the system
in its spin-$N/2$ incarnation \rfs{eq:6}. Translated to the language
of this Hamiltonian a sparse initial occupation of fermionic Hilbert
space means that we start the process at zero bosons and a spin that
is pointing nearly downwards (while the total spin continues to define
the dimension of state space, $S=N/2$.)  The particle conservation law
then reads $S_z + n = -N/2+\epsilon$, where $\epsilon \ll N$ and $0
\le n \le \epsilon$ is the number of bosons. This regime can be
modeled in terms of Holstein-Primakoff bosons. However, as the spin
continues to move downwards, the regime of Holstein-Primakoff
linearizability will not be left.  Introducing
\begin{align}
  \label{eq:8a}
  S^+&= c^\dagger (2S-c^\dagger c)^{1/2},\nonumber\\
  S^-&=(2S-c^\dagger c)^{1/2} c ,\nonumber\\
  S^z&=-S+c^\dagger c,
\end{align}
where $S=N/2$ (the version of \rfs{eq:8} for the spin pointing
close to downwards), we find
$$H = \lambda t \left( b^\dagger b -
  c^\dagger c \right) + g \left( b^\dagger c + c^\dagger b \right).
$$
Here $b^\dagger b+ c^\dagger c = \epsilon$. This problem is equivalent
to the standard Landau-Zener problem. This proves that in this sector
the model reduces to the standard problem and does not have any novel
features.

\section{Conclusions}
\label{sec:conclusions}

Let us summarize the findings of our paper. The model described by
\rfs{eq:6}, (as well as its other incarnations, given by
Eqs.~\rf{eq:5}, \rf{eq:4}, and \rf{eq:17} together with
\rf{eq:hamdifbos}), with the initial conditions
$$ \lim_{t \rightarrow -\infty} \left< b^\dagger(t) b(t) \right> =0,$$
results in a total final number of bosons $n_b$, defined by
$$ n_b = \lim_{t \rightarrow +\infty}  \left< b^\dagger(t) b(t) \right>
$$
given by the following expressions:
\begin{itemize}
\item For small values of the LZ-parameter,
$$
e^{\frac {\pi g^2}{\lambda}} \ll N:\qquad  n_b = e^{\frac {\pi g^2}{\lambda}}-1.
$$
This result can be obtained within the Holstein-Primakoff method, or by the large-$N$ approximation within the Keldysh formalism.

\item At intermediate values,
$$
e^{\frac {\pi g^2}{\lambda}} \simeq N:\qquad  n_b \simeq \frac{N \left( e^{\frac {\pi g^2}{\lambda}}- 1\right)}{2 e^{\frac {\pi g^2}{\lambda}} + N},
$$
This can be found using the kinetic equation approach within the Keldysh formalism.

\item And at large values,

$$
e^{\frac {\pi g^2}{\lambda}} \gg N:\qquad n_b \simeq N \left( 1-\frac{\lambda}{\pi g^2} \log(N) \right),
$$
This result was found using the quasiclassical approximation together with the adiabatic invariants formalism.
\end{itemize}
We observe main distinctions between our results and what would
have happened had we extrapolated the standard Landau-Zener
to our problem (which would have given $n_b = N ( 1-e^{-\frac{\pi
      g^2}{\lambda}} )$.)  First, the transition between
the fast rate to slow rate regime happens at
$$ \lambda \simeq \frac{\pi g^2}{\log N},$$
the rate which becomes progressively smaller as $N$ is increased
(unlike the Landau-Zener result $\lambda \simeq \pi g^2$).  Second, if
$\lambda \ll \pi g^2/\log N$, then $n_b$ approaches $N$ as a linear
function of $\lambda$, much more slowly than the Landau-Zener result
which would predict an exponentially fast approach. Finally, particle
occupancies in our system show massive quantum fluctuations,
comparable in value to the mean particle numbers.

\textit{Acknowledgments:} A.A. acknowledges discussions with Fritz
Haake. Work supported by SFB/TR 12 of the Deutsche Forschungsgemeinschaft, by the NSF grant DMR-0449521, and by AFOSR YIP.

\appendix

\section{Derivation of Eqs. (\ref{eq:15}) and (\ref{eq:19})}
\label{sec:derivation-eqs.-xx}

Consider the Heisenberg equations of motion $i\partial_t x = [\hat
H,x],\;   x=c,b$ corresponding to the Hamiltonian (\ref{eq:9}),
\begin{align}
  \label{eq:end}
  i\partial_t b + \lambda t b + g c^\dagger &=0,\nonumber\\
  -i\partial_t c^\dagger + \lambda t c^\dagger + g b &=0.
\end{align}
The structure of these equations suggests the linear combination
\begin{align}
\label{eq:11}
  c &= \mu c_0+ \nu b_0^\dagger\nonumber\\
  b &= \rho b_0+  \sigma c_0^\dagger,
\end{align}
where $c_0\equiv c(t_0)$, $t_0\ll 0$ is some time in the distant past
when the dynamical evolution is started, and $\mu,\dots,\sigma$ are
complex valued functions depending on time. Differentiating the first
of the two equations in (\ref{eq:end}) once more w.r.t. time and
substituting the second equation, we obtain
\begin{align}
  \label{eq:10}
  \left[\partial_t^2 -i\lambda  + (\lambda t)^2-g^2\right]x=0,\;\;x=c,b.
\end{align}
These two equations are coupled by the initial conditions, which
follow from the first order ancestor equations (\ref{eq:end}). To make
these dependencies explicit, we substitute the expansion (\ref{eq:11})
and obtain
\begin{align}
  \label{eq:12}
& \left[\partial_t^2 -i\lambda  + (\lambda
    t)^2-g^2\right]\kappa=0,\;\;\kappa=\mu,\nu,\rho,\sigma,\nonumber\\
&\nonumber\\
&\kappa=\mu,\rho:\;\;\kappa(t_0)=1,\qquad\partial_t \kappa(t_0) = i \lambda t_0,\nonumber\\
&\kappa=\nu,\sigma:\;\;\kappa(t_0)=0,\qquad\partial_t \kappa(t_0) = i g.
\end{align}
Notice that
\begin{align}
  \label{eq:13}
&  \mu = \rho \;\mathrm{and}\;\nu=\sigma,\nonumber\\
&  |\mu|^2 - |\nu|^2=1,
\end{align}
where the first line is a consequence of $(\mu,\rho)$ and
$(\nu,\sigma)$ obeying differential equations with identical initial
conditions, and the second is enforced by the fact that the time dependent operators
$(c,c^\dagger,b,b^\dagger)$ obey canonical commutation relations.

These equations afford a solution in terms of parabolic cylinder
functions~\cite{Kayali2003}. All we need to know about these functions
is that
\begin{equation}
  \label{eq:14}
  |\mu|^2\stackrel{t\to \infty}{\longrightarrow} e^{\pi g^2/\lambda}.
\end{equation}
Using $n_b = \langle c^\dagger c\rangle = \langle (\bar \mu
c_0^\dagger +\bar \nu b_0)(\mu c_0+ \nu b_0^\dagger)\rangle =
|\nu|^2$ and Eqs.~(\ref{eq:13}), (\ref{eq:14}) we then obtain Eq. (\ref{eq:15}).

The actual distribution of the particle number can be obtained by a
bit of linear algebra. We begin by condensing Eqs.~(\ref{eq:11}) into
the matrix equation
\begin{equation}
  \label{eq:16}
  \left(
    \begin{matrix}
      c\cr
      b^\dagger
    \end{matrix}
\right)
=
\left(
  \begin{matrix}
    \mu & \nu\cr
    \bar \nu & \bar \mu
  \end{matrix}
\right)
  \left(
    \begin{matrix}
      c_0\cr
      b_0^\dagger
    \end{matrix}
\right)\Rightarrow
  \left(
    \begin{matrix}
      c_0\cr
      b_0^\dagger
    \end{matrix}
\right)
=
\left(
  \begin{matrix}
    \bar \mu & -\nu\cr
    -\bar \nu & \mu
  \end{matrix}
\right)
  \left(
    \begin{matrix}
      c\cr
      b^\dagger
    \end{matrix}
\right),
\end{equation}
where we used Eqs.~(\ref{eq:13}).
Denoting the occupation number eigenstates of the time dependent
operators by $|k,l\rangle$, i.e.
\begin{align*}
  c|k,l\rangle = \sqrt k |k-1,l\rangle,\qquad &c^\dagger |k,l\rangle =
  \sqrt{k+1} |k+1,l\rangle,\\
  b|k,l\rangle = \sqrt l |k,l-1\rangle,\qquad &b^\dagger |k,l\rangle =
  \sqrt{l+1} |k,l+1\rangle,
\end{align*}
We now expand the vacuum state (of the operators $c_0,b_0$),
$|0\rangle$, in the basis $\{|k,l\rangle \}$:
$$
|0\rangle = \sum_{k,l} c_{k,l} |k,l\rangle.
$$
To determine the coefficients $c_{k,l}$, we use the defining relation
$0\stackrel{!}{=}c_0|0\rangle \stackrel{(\ref{eq:16})}{=}(\bar \mu c
- \nu b^\dagger )\sum_{k,l} c_{k,l} |k,l\rangle$. This equation, and
its partner $b_0|0\rangle=0$ generate the set of recursion relations
\begin{align*}
  &c_0|0\rangle=0:\;&\mu  \sqrt{k+1} c_{k+1,l+1}=\bar \nu \sqrt{l+1}
  c_{k,l},&\;\;\;k,l\ge 0,\\
&&\mu c_{k,0}=0,&\;\;\;k\ge 1,\\
&&&\\
   &b_0|0\rangle=0:\;&\bar \nu  \sqrt{k+1} c_{k,l}=\mu \sqrt{l+1}
  c_{k+1,l+1},&\;\;\;k,l\ge 0,\\
&&\mu c_{0,l}=0,&\;\;\;l\ge 1.
\end{align*}
 From these equations we deduce $c_{k,0}=0$ for $k\ge 0$ and
 $c_{0,l}=0$ for $l\ge 0$. The first and third equation then imply
 $c_{k,l}=0$ for $k\not=l$. The diagonal terms successively descend
 from $c_{0,0}$ according to
$$
c_{k,l} = \delta_{k,l}\, c_{0,0} \left({\bar \nu \over \mu}\right)^k.
$$
The normalization condition $1=\langle 0|0\rangle = \sum_{k,l}
|c_{k,l}|^2$ generates the additional condition
$$
1
= {|c_{0,0}|^2\over {1- \left|{\bar \nu \over \mu}\right|^{2}}}=
  |c_{0,0}|^2 |\mu|^2.
$$
We thus arrive at the expansion
$$
|0\rangle = {1\over |\mu|}\sum_{k=0}^\infty \left({\bar \nu \over
    \mu}\right)^k |k,k\rangle.
$$
From this, the moments of the particle number operator are
straightforwardly obtained as
$$
c_m \equiv \langle 0|(c^\dagger c)^m|0\rangle =
{1\over |\mu|^2} \sum_{k=0}^\infty {| \nu|^{2k} \over
    |\mu|^{2k}} k^m.
$$
Comparison with the formulation of moments in terms of the discrete probability
distribution $P(n)$
$$
c_m = \sum_n P(n) n^m
$$
leads to the identification $P(n) = {1\over |\mu|^2} {| \nu|^{2n} \over
    |\mu|^{2n}}$ or, using Eq.~(\ref{eq:13}),
\begin{equation}
  \label{eq:18}
  P(n) = {1\over |\mu|^2}\left(1-{1\over |\mu|^2}\right)^n.
\end{equation}
Substitution of Eq.~(\ref{eq:14}) then leads to
Eq.~(\ref{eq:19}). From this distribution, the mean value is readily
obtained as (\ref{eq:15}).

\section{Derivation of the Truncated Wigner Approximation for the spin-boson model.}
\label{app:twa}

\subsection{Truncated Wigner approximation}
\label{sec:trunc-wign-appr}

In this appendix, we first review the TWA for a general system of
bosons, and then adapt to our specific spin-boson problem. Consider a
(generally time dependent) Hamiltonian formulated in terms of normal
ordered bosonic creation and annihilation operators $\mathcal
H(\hat\psi^\dagger,\hat\psi,t)$. The hat-notation $\hat\psi$ is used
to distinguish operators from $c$-numbers. In the classical limit the
operators $\hat\psi$ are substituted by complex numbers $\psi$
satisfying the Hamiltonian equation of motion (Gross-Pitaevskii
equation):
$$
i\hbar{\partial \psi\over\partial t}={\delta \mathcal H(\psi^\ast,\psi,t)\over\delta \psi^\ast},
$$
where the right hand side denotes a variational
derivative. Classically,  these
equations have to be supplied by definite initial
conditions $(\psi_0,\psi_0^\ast)$.  Within the truncated Wigner
approximation -- the first order quantum correction to the classical
picture -- the initial data becomes a distribution
$W(\psi_0,\psi_0^\ast) \,d\psi_0 d\psi_0^\ast$. The kernel of this
distribution is defined by
\begin{align}
\label{wig_coherent}
   W_0(\psi_0,\psi_0^\ast)&=\int d\eta_0^\ast d\eta_0 \langle
\psi_0-{\eta_0\over 2}|\rho | \psi_0+{\eta_0\over
2}\rangle\cr
&\times\mathrm e^{-|\psi_0|^2-{1\over
4}|\eta_0|^2}\,\mathrm e^{{1\over
2}(\eta_0^\ast\psi_0-\eta_0\psi_0^\ast)},
\end{align}
where $\rho$ is the initial density matrix of the system. Expectation
values of operators $\Omega(\hat\psi^\dagger,\hat\psi)$ at time $t$ are then to be calculated as~\cite{ap_twa}
\begin{equation}
  \langle
\Omega(t)\rangle=\int\int d \psi_0 d \psi_0^\ast\,
W_0(\psi_0,\psi_0^\ast)\Omega_{cl}(\psi(t),\psi^\ast(t)),
\label{eq:omega5},
\end{equation}
where $\psi(t)$ is the solution of the classical Gross-Pitaevskii
equations of motion with initial condition $\psi_0$ and
$\Omega_{cl}(\psi,\psi^\ast)$ is the Weyl symbol of the
operator $\Omega$. For a normal ordered  $\Omega$ the latter is
defined by
\begin{equation}
  \Omega_{cl}(\psi,\psi^\ast)=\int\int d\eta
d\eta^\ast\, \Omega\left(\psi-{\eta/
    2},\psi^\ast+{\eta^\ast/ 2}\right)\mathrm e^{-|\eta|^2/2},
\label{Weyl_coherent}
\end{equation}
where $\Omega(\psi,\psi^\ast)$ is obtained by substitution of operators in
$\Omega(\hat \psi,\hat \psi^\dagger)$ as $\hat \psi\to \psi$ and $\hat
\psi^\dagger \to \psi^\ast$. For example for the number operator $\Omega=\hat\psi^\dagger \hat\psi$ we get
\begin{equation}
\Omega_{cl}=\int\int d\eta
d\eta^\ast\, (\psi-{\eta/
    2})(\psi^\ast+{\eta^\ast/ 2})\mathrm e^{-|\eta|^2/2}=\psi^{\ast}\psi-{1\over 2}.
\end{equation}
The same result can be obtained by first symmetrizing number operator with respect to $\hat\psi^\dagger$ and $\hat\psi$:
\begin{equation}
{\hat\psi^\dagger\hat\psi}={\hat\psi^\dagger\hat\psi+\hat\psi\hat\psi^\dagger\over  2}-{1\over 2}
\end{equation}
and then substituting the operators $\hat\psi^\dagger$ and $\hat\psi$ with the numbers $\psi^\ast$ and $\psi$ (see Ref.~\cite{blakie}) for more details.

% {\tt what is meant by this?}  Alternatively the Weyl symbol is given
% by the symmetrized for of the operator $\Omega$ with respect to
% $\hat\psi$ and $\hat\psi^\dagger$.

\subsection{Adaption to the spin-boson problem}
\label{sec:adaption-spin-boson}

Our next task is to generalize the TWA to the Hamiltonian
(\ref{eq:6}). To this end, we employ the
Schwinger representation of spins through bosons $\hat\alpha$ and
$\hat\beta$:
\begin{equation}
  \hat{S}^z={\hat{\alpha}^\dagger\hat{\alpha}-\hat{\beta}^\dagger\hat{\beta}\over
  2},\; \hat{S}^{+}=\hat{\alpha}^\dagger\hat{\beta},\;
\hat{S}^-=\hat{\beta}^\dagger\hat{\alpha}.
\label{schwing}
\end{equation}
Here, the boson operators $\alpha$ and $\beta$ satisfy the additional
constraint $\hat{n}=\hat{\alpha}^\dagger \hat{\alpha}+
\hat{\beta}^\dagger\hat{\beta}=2S$. The operator $\hat n$
commutes with all spin operators which means that the fulfilment of the
constraint at an arbitrary time will be conserved for all later
times. The option of a purely bosonic representation means that the
TWA can readily be generalized to the  Hamiltonian~(\ref{eq:6}). Once
the TWA has been formulated for the Schwinger bosons, we are at
liberty to switch back to spin variables
Eqs.~(\ref{schwing}). However, at this point, the spins have become
classical numbers, defined in terms of the $c$-number valued
boson variables $\alpha$ and $\beta$.

The classical equations of motion are given by Eq.~(\ref{cl_eq4}).
The initial density matrix describes a pure spin state, polarized
along the $z$-axis or,  in bosonic language, a state with $2S$
$\alpha$ bosons and no $\beta$ bosons. (For
the spin pointing along $-z$ $\alpha$ and $\beta$ should be
interchanged). The corresponding Wigner function then
reads~\cite{walls-milburn, ap_twa, ap_cat}: \be
W(\alpha,\alpha^\ast,\beta, \beta^\ast)=2\mathrm
e^{-2|\alpha|^2-2|\beta|^2} L_{2S}(4|\alpha|^2),
\label{wig_spin}
\ee where $L_N(x)$ is the Laguerre's polynomial of order $N$. At large
$S$, the latter is strongly oscillatory and the Wigner transform is
localized near $|\alpha|^2=2S+1/2$ (see Ref.~\cite{ap_twa}). So in
this case to a very good accuracy (up to $1/S^2$) one can use
\begin{equation}
  W(\alpha,\alpha^\ast,\beta,
\beta^\ast)\approx \sqrt{2}\mathrm e^{-2|\beta|^2}
\delta(|\alpha|^2-2S-1/2).
\label{wig_spin_1}
\end{equation}
If we reexpress this distribution function through spins then to the same accuracy we will find
$$
W(S_z,S_\perp)\approx {1\over \pi S}\mathrm
e^{-S_\perp^2/S}\delta(S_z-S),
$$
where $S_\perp^2=S_x^2+S_y^2$. This Wigner function has a transparent
interpretation. If the quantum spin points along the $z$ direction,
because of the uncertainty principle, the transverse spin components
still experience quantum fluctuations so that
$$
\langle S_x^2\rangle=\langle S_y^2\rangle={S\over 2},
$$
which is indeed the correct quantum-mechanical result. Finally, the
distribution of the $b$-boson  represents the Gaussian Wigner function
of the vacuum state,
$$
W(b,b^\ast)=2\exp[-2b^\ast b]\equiv W(n)=2\exp[-2n],
$$
where in the second representation we switched to a representation in
terms of the boson number $n=b^\ast b$.

\subsection{TWA vs quantum solution of linear regime}
\label{sec:twa-vs-quantum}

We here discuss how the solution of the equations of motion (\ref{linear_semicl}) obtains information
equivalent to that of the full quantum solution of the linear regime. To this end, we
notice that the equations afford a solution as (see Eq.~(\ref{eq:11}))
$$
b(t)=\mu b(t_0)+\nu s^-(t_0),\quad s^-(t)=\mu^\ast s^-(t_0)+\nu^\ast b(t_0),
$$
where at $t\to\infty$ we have $|\mu|^2=x$ and $|\nu^2|=x-1$ (see Eqs.~(\ref{eq:13}) and (\ref{eq:14}). If we are interested only in the statistics of the number of bosons we do not need to know the
phases of $\mu$ and $\nu$. Using the Wigner function (\ref{wig_coherent}) to compute the average and the Weyl symbol~(\ref{weyl}) for the operators $n_b$, $n_b^2$ it is then straightforward to obtain
$$
\langle n_b\rangle =x \langle n_b(t_0) \rangle +(x-1)\langle
|s^-(t_0)|^2\rangle -{1\over 2}=x -1,
$$
and the second moment
\begin{align*}
&\langle n_b^2\rangle=-x+{1\over 2}+x^2\langle n_b^2(t_0)\rangle+(x-1)^2\langle |s^-(t_0)|^4\rangle\\
&+4x(x-1) \langle n_b(t_0)\rangle\langle
|s^-(t_0)|^2\rangle=2x^2-3x+1.
\end{align*}
It is easy to check that this
result conforms with the distribution (\ref{eq:19}).

\section{The meaning of the fixed points}
\label{sec:timeind}
Let us consider the time independent version of the model studied in this paper. It is given by
the time independent version of the Hamiltonian \rfs{eq:6}
\be  \hat H = -  \frac{\gamma}2  ~ b^\dagger  b + \frac{\gamma}2 S^z +
\frac{g}{\sqrt{N}}  \left(  b^\dagger \,
  S^- +  b \, S^+ \right).
\end{equation}
In the sector where $$b^\dagger b + S^z = \frac{N}2,$$
which is the one mostly studied in this paper, this Hamiltonian can be thought of as simply a $N+1$ by $N+1$ matrix. In the basis
of states $\left| n \right>$ where $n$ denotes the boson occupation number, this matrix takes the form
\begin{eqnarray} \label{eq:mmatt} H_{n',n} &=&  -  \gamma n \, \delta_{n',n} + \frac{g}{\sqrt{N}} n \sqrt{N-n'} \, \delta_{n'+1,n} +\cr
&&   \frac{g}{\sqrt{N}} n' \sqrt{N-n} \, \delta_{n'-1,n}
\end{eqnarray}
(compare with  \rfs{eq:matrixelement}).

It is easy to evaluate the eigenvalues of this matrix numerically for moderate $N$.

Now consider a semiclassical version of this Hamiltonian, \rfs{eq:43}, which for completeness we reproduce here
$$ H =- \gamma n + 2 n \sqrt{1-n} \cos(\phi). $$
Here both the Hamiltonian and the variable $n$ were rescaled according to Eqs.~\rf{eq:rescale} and \rf{eq:rescaleH} to ease
comparison with the language of section \ref{sec:deepadiabaticlimit}, although we need to keep that in mind when comparing it with
\rfs{eq:mmatt} which was not rescaled in any way.
Let us minimize this Hamiltonian with respect to $n$ and $\phi$. Its minimum is given by the substitution of the fixed point $\phi=0$, and $n_1(\gamma)$ given by $n_1(\gamma) =0$ for $\gamma<-2$ and by \rfs{eq:47a}, or
$$n_1(\gamma) = \frac{12-\gamma^2+\gamma \sqrt{12+\gamma^2}}{18}
$$
 if $\gamma\ge-2$. To find the energy minimum we substitute $n_1(\gamma)$ into $H$ to find
\begin{eqnarray}  \label{eq:classminn} E_{\rm min} =N  \min_{n,\phi} H = 0,  &{\rm if} & \gamma < -2 \cr
E_{\rm min} =   - N \, \frac{36 \gamma - \gamma^3 + (12+\gamma^2)^{\frac 3 2}}{54} & {\rm if} &  \gamma \ge -2.
\end{eqnarray}

\begin{figure}[h]
  \centering
  \includegraphics[width=3.3in]{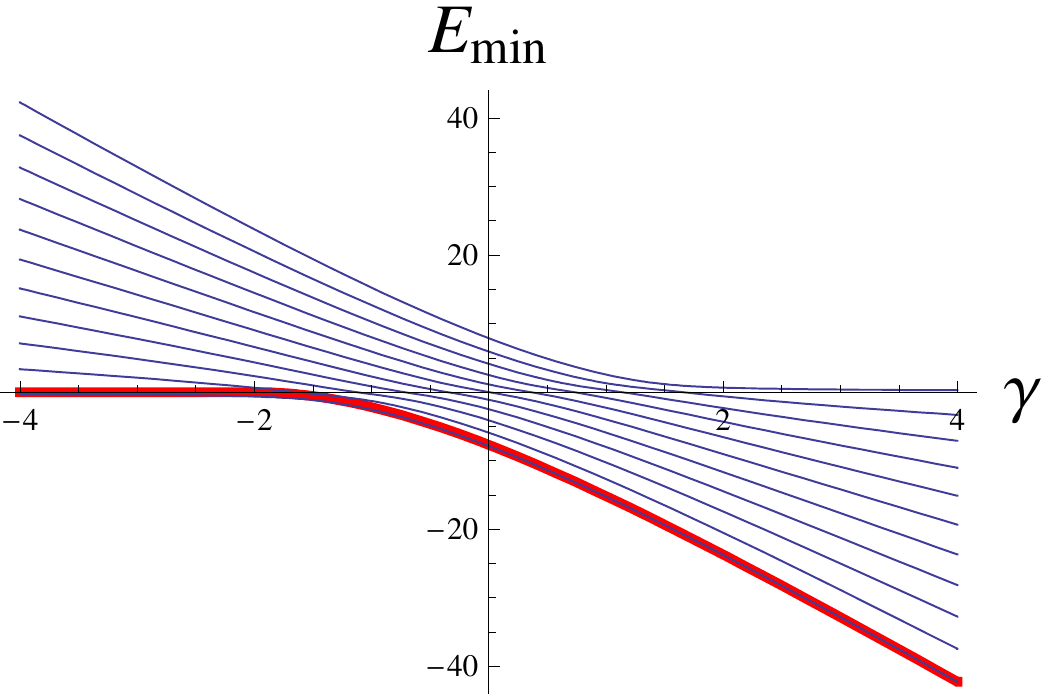}
  \caption{\label{fig:qpt}
 The energy levels (eigennvalues of the matrix \rfs{eq:mmatt}) for $N=10$ are plotted as a function of $\gamma$. At the same time, the minimum of the
 classical Hamiltonian, given by \rfs{eq:classminn}, is also plotted (red curve). We see that \rfs{eq:classminn} closely follows the ground state energy
 of \rfs{eq:mmatt}. }
\end{figure}

The minimum of the classical Hamiltonian \rfs{eq:classminn}, together with the eigenvalues of \rfs{eq:mmatt}, are plotted as a function
of $\gamma$ on Fig.~\ref{fig:qpt} for $N=10$. One can see that the classical minimum closely follows the quantum ground state.
Yet \rfs{eq:classminn} cannot be the exact ground state of the problem at this value of $N$ since it is not analytic in $\gamma$. However, it can be
the exact ground state in the limit $N \rightarrow \infty$.
Thus we conjecture that \rfs{eq:classminn} {\sl is} the ground state of the Hamiltonian in our problem
in the limit $N \rightarrow \infty$.     If so, it implies that our problem undergoes a quantum phase transition as a function of $\gamma$.

When $\gamma$ depends on time, it follows that we drive our system across the quantum phase transition. The existence of the critical trajectory
and the singular behavior of the adiabatic invariants discussed in section \ref{sec:deepadiabaticlimit} can be traced to the existence of this transition.
The transition in the time independent version of the Dicke model is well known in the literature (see, for example, Ref.~\cite{Dimer2007} and references therein).

\section{Derivation of Eq.~(\ref{eq:29})}
\label{sec:epsilon}

Here, just as in the section~\ref{sec:deepadiabaticlimit}, we use the rescaled variables $n$ and $H$, according to Eqs.~\rf{eq:rescale}, \rf{eq:rescaleH}. 
To obtain the value of $\epsilon=\epsilon^\ast$ at which $w=\pi$, we
need to inspect the increment of the angular variable
$w$
in the problem where $\gamma$ changes in time.  It satisfies the
equation \cite{LL1}
$$
d_t w = \omega + \left( \frac{\partial
    \Lambda}{\partial I} \right)_{w,\gamma} \lambda,
$$
where $\Lambda$ was defined in \rfs{eq:26}. At small $\lambda$ we
neglect the second term to arrive at
\begin{equation}
\label{eq:wofepsilon} w(\epsilon)
\approx \int_{\epsilon_0}^{\epsilon} d\epsilon' \frac{ \pi
  \sqrt{\epsilon'}}{ \lambda \log \left[-\frac{16 {\epsilon'}^2}{H}
  \right]}.
\end{equation}
Here we used Eq.~\eqref{eq:28}, the identity $d\epsilon/dt = 2
\lambda$, and $\epsilon_0>0$ denotes the moment of time when the
system crosses the critical trajectory and switches to trajectories
winding about the critical point \rfs{eq:48}. This allows us to find
$w$ as a function of $\epsilon$, which in turn represents time.  To
compute this integral, we need to know $H(\epsilon)$. This in turn can
be estimated by noting that
$$
d_t H = -2 \lambda n.
$$
Here, $n$ itself is a function of time, which satisfies its equation
of motion $ d_t n = - 2 n \phi$.  Using \rfs{eq:condonfi} we find
\begin{equation}
  \label{eq:nn}
  n = n_0 e^{-\frac 1 \lambda \int_{\epsilon_0}^{\epsilon} d\epsilon' \phi}  = n_0 e^{\frac 2{3\lambda} \left( \epsilon^{\frac 3 2}- \epsilon_0^{\frac 3 2} \right)},
\end{equation}
which integrates to
\begin{equation}
  \label{eq:HH}
  H = - n_0 \int_{\epsilon_0}^\epsilon d\epsilon'   e^{\frac 2{3\lambda} \left( {\epsilon'}^{\frac 3 2}- \epsilon_0^{\frac 3 2} \right)}.
\end{equation}
This needs to be substituted into \rfs{eq:wofepsilon}. We are now in a
position to estimate the value $\epsilon^\ast$ corresponding to
$w=\pi$.
To compute the integrals in \rfs{eq:wofepsilon}, we need to study the behavior of $H$. At small $\epsilon-\epsilon_0$, $H$
can be approximated as
$$
H \approx - n_0 \left(\epsilon-\epsilon_0 \right).
$$
At larger $\epsilon$, we can estimate it by
$$
H  \approx -n_0 \frac{\lambda}{\sqrt{\epsilon}} e^{\frac{2}{3\lambda} \epsilon^{\frac 3 2}}.
$$
Here, $\epsilon_0$ represents the moment in time when the system
crosses the critical trajectory. At this point $n$ did not have a
chance to change appreciably from its initial value, $n( \gamma
\rightarrow -\infty)$. Thus $n_0 \sim \frac{1}{N}$. This means that to
logarithmic accuracy, we may use the approximation $H(\epsilon)\sim
-1/N$ in \eqref{eq:wofepsilon}.  Other factors, such as $16
\epsilon^2$ under the logarithm can also be neglected. This leads to
$$
w =
\int_{\epsilon_0}^{\epsilon^*} d\epsilon \frac{\pi
  \sqrt{\epsilon}}{\lambda \log \left[ N
      \right]} \sim \frac{2
  {\epsilon^*}^{\frac 3 2}}{3 \lambda \log N}  \stackrel{!}{\sim} \pi.
$$
Solution of the this equation for $\epsilon^\ast$ obtains the result
\eqref{eq:29}. Our derivation has been self-consistent in the sense
that even at the maximal value $\epsilon^\ast$,
$$
\left| H(\epsilon^*) \right| \sim
\frac{\lambda^{\frac 1 3}}{\left( \frac 3 2 \log N \right)^{\frac 2
    3}} \ll {\epsilon^*}^2,
$$
where again the conditions \rfs{eq:limit} were used.

The last thing which remains to be done is to check that
\rfs{eq:condonfi} is consistent with the behavior of the system.  The
equation of motion \rfs{eq:eqonfi}, together with the condition that
$\left| H \right| \ll \epsilon^2$, implies
$$
2\lambda \, d_\epsilon
\phi = -\epsilon+\phi^2,
$$
assuming that $\phi^2 \le \epsilon$.  This
is a Riccati equation, which can be brought to the form of the
Schr\"odinger equation by the substitution
$$
R=e^{-\frac 1 {2
    \lambda} \int \phi d\epsilon},
$$ to give
$$ -2 \lambda
\frac{d^2 R}{d \epsilon^2} + \epsilon R = 0.
$$
This is the equation for the Airy function. It can be investigated
using the WKB approximation, which reproduces the ansatz
$\phi=-\sqrt{\epsilon}$.  Close to $\epsilon=0$, the WKB
approximation breaks down, so this ansatz is no longer
correct. However, it is easy to modify relations such as \rfs{eq:nn}
and \rfs{eq:HH} by substituting $R^2$, with $R$ being the appropriate
Airy function, in place of the exponentials in these relations. This
does not affect the final answer for $\epsilon^*$, and consequently
for $I_{final}$.

\bibliography{paperLZV}

\end{document}